%% file: paper.tex
\documentclass[oldversion]{aa}
\usepackage{graphicx}
\usepackage{txfonts}
\usepackage{booktabs}
\usepackage{natbib}
\usepackage[utf8]{inputenc}
\usepackage{multirow}
\bibpunct{(}{)}{;}{a}{}{,} 

\newcommand{\swift}{\textit{Swift}}

\newcommand{\fermi}{\textit{Fermi}}

\usepackage[usenames,dvipsnames]{color}
\usepackage{colortbl,xcolor}

\begin{document}

\title{Multi-color observations of short GRB afterglows: \\
20 events observed between 2007 and 2010}

\author{
A.~Nicuesa Guelbenzu\inst{1}, 
S. Klose\inst{1}, 
J. Greiner\inst{2}, 
D. A. Kann\inst{1}, 
T. Kr\"uhler\inst{3}, 
A. Rossi\inst{1}, 
S. Schulze\inst{4},
P.~M.~J.~Afonso\inst{5},
J.~Elliott\inst{2},
R.~Filgas\inst{2,6},
D. H. Hartmann\inst{7},
A.~K\"upc\"u~Yolda\c{s}\inst{8},
S. McBreen\inst{9},
M.~Nardini\inst{10},
F.~Olivares~E.\inst{2},
A.~Rau\inst{2},
S.~Schmidl\inst{1},
P.~Schady\inst{2},
V.~Sudilovsky\inst{2},
A.~C.~Updike\inst{11}, \&
A.~Yolda\c{s}\inst{8}
}

\offprints{A. Nicuesa Guelbenzu, ana@tls-tautenburg.de}

\institute{Th\"uringer Landessternwarte Tautenburg, Sternwarte 5, 07778 Tautenburg, Germany\\
\email{ana@tls-tautenburg.de}
\and
Max-Planck-Institut f\"ur Extraterrestrische Physik, Giessenbachstra\ss{}e,
85748 Garching, Germany
\and
Dark Cosmology Centre, Niels Bohr Institute, University of
Copenhagen, Juliane Maries Vej 30, 2100 Copenhagen, Denmark
\and
Centre for Astrophysics and Cosmology, Science Institute, University of Iceland, Dunhagi 5, 107 Reykjav\'ik, Iceland 
\and
American River College, Department of Physics and Astronomy, 4700 College Oak Drive,
Sacramento, CA 95841, USA
\and
Institute of Experimental and Applied Physics, Czech Technical University,
Horsk\'a 3a/22, 12800 Prague, Czech Republic 
\and
Clemson University, Department of Physics and Astronomy, Clemson, SC
29634-0978, USA
\and
Institute of Astronomy, University of Cambridge, Madingley Road CB3 0HA, Cambridge, UK
\and
School of Physics, University College Dublin, Dublin 4, Republic of Ireland
\and
Universit\`a degli studi di Milano-Bicocca, Piazza
della Scienza 3, 20126 Milano, Italy
\and
Department of Astronomy, University of Maryland, College Park, MD 20742, USA 
}

\date{Received 2012 May 4; accepted XXXX}

\authorrunning{Nicuesa Guelbenzu et al.}
\titlerunning{short GRBs}

\abstract{   
We report on follow-up observations of 20 short-duration gamma-ray bursts
(GRBs; $T_{90}<2$~s)  performed in $g'r'i'z'JHK_s$ with the seven-channel
imager GROND between mid-2007 and the end of 2010. This is one of the most
comprehensive data sets on GRB afterglow observations of short bursts
published so far. In three cases GROND was on target within less than 10 min
after the trigger, leading to the discovery of the afterglow of GRB 081226A
and its faint underlying host galaxy. In addition, GROND was able to image the
optical afterglow and follow the light-curve evolution in further five cases,
GRBs 090305, 090426, 090510, 090927, and 100117A.  In all other cases
optical/NIR upper limits can be provided on the afterglow magnitudes. 
After shifting all light curves to a common redshift we find that the optical
luminosities of the six events with light curves group into two subsamples.
GRBs 090426 and 090927 are situated in the regime occupied by
long-duration events (collapsars),  while the other four bursts occupy the
parameter space typical for merger events, confirming that
the short-burst population is contaminated by collapsar events.  Three of the
aforementioned six bursts with optical light curves show a break.  In addition
to GRBs 090426 and 090510 (paper I, II), also for  GRB 090305 a break is
discovered in the optical bands at 6.5 ks after the trigger. For GRB 090927 no
break is seen in the optical/X-ray light  curve until about 150~ks/600~ks
after the burst. The GROND multi-color data support the view that this
burst is related to a collapsar event. For GRB 100117A a decay slope of its
optical afterglow could be measured.  For all six GRBs
at least a lower limit on the corresponding jet opening angle can be
set. Using these data, supplemented by a about 10  events taken from the
literature, we compare the jet half-opening angles of long and short bursts.
We find tentative evidence that short bursts have  wider opening angles than
long bursts. However, the statistics is still very poor.}

\keywords{Gamma rays: bursts}

\maketitle

\section{Introduction}\label{Intro}

Gamma-Ray Bursts (GRBs) show a bimodality in  their duration distribution,
separated in the CGRO/BATSE data at $T_{90}=2$~s, with the peak of the
short-burst population at $T_{90}\sim$0.5~s and the long-burst population at
$\sim$30~s (\citealt{Kouveliotou1993,Sakamoto2011ApJS}).  Historically,
bursts are still devided into long and short based on the BATSE scheme, even
though the shape of the bimodal distribution is energy-dependent, in
particular peaking for \swift/BAT at $T_{90}\sim0.5$~s and $\sim70$~s,
respectively  (\citealt{Sakamoto2011ApJS}).
 
According to the current picture, long bursts originate from the collapse of
massive stars into black holes (\citealt{MacFadyen1999}) or into rapidly
spinning, strongly magnetized neutron stars (e.g.,
\citealt{Usov1992,Mazzali2006}).  Short bursts are instead commonly attributed
to the merger of compact stellar  objects (e.g.,
\citealt{Paczynski1986,Nakar2007}). The physical association of long bursts
with the collapse of massive stars has been well established (e.g.,
\citealt{ZehKH2004,Hjorth2003Natur,Pian2006Natur,Ferrero2006,WB2006,
  Fruchter2006}). However, the observational situation with short bursts is
less clear. 

Until 2005 no afterglow of a short burst had ever been detected, while for the
long burst sample  at that time many important discoveries had already
been made (redshifts, supernova light, collimated explosions, circumburst wind
profiles). The first well-localized short burst (GRB 050509B;
\citealt{Gehrels2005Natur}) was seen close  in projection to a massive
early-type galaxy (\citealt{Hjorth2005,Bloom2006ApJ638}), supporting  the
model that compact stellar mergers are the progenitors of short-duration
gamma-ray bursts. However, since then the  observational progress has been
rather modest when compared to the long-burst population (for a review
\citealt{Gehrels2009,Berger2011NewAR}). 

There are mainly two reasons for this situation. Firstly,  compared to long
bursts there is a substantially smaller detection rate of short bursts.
Secondly, short-burst afterglows are rarely brighter than  $R=20$ even minutes
after a trigger (e.g., \citealt{Kann2010,Kann2011}).  This general faintness
makes their discovery and detailed follow-up very challenging.  However, only
the precise detection of the afterglow, with sub-arcsec accuracy, enables a
secure determination of a putative GRB host  galaxy and its redshift, while
the  X-ray plus optical light curves provide  information about the processes
that take place after the explosion, clues about the physics of the central
engine, and the properties of the environment of the progenitor. Rapid
follow-up observations of these events are therefore very important to gain as
much observational data as possible.

Since there is a substantial overlap between the long and the short-burst
duration distribution, the simple devision between long and short is only a
first guess about the true origin of a burst under consideration.  Several
other phenomenological properties of the bursts and their afterglows have to be
considered in order to reveal the nature of their progenitors
(\citealt{Zhang2007,Zhang2009,Kann2011}). Thereby, of special interest are the
circumburst density profiles, the  afterglow luminosities,   and the outflow
characteristics that might be shaped  by or related to the physical properties 
of the GRB progenitors.

Theoretical studies suggest that long GRBs are followed by more luminous
afterglows than short bursts, mainly due to the expected difference in the
circumburst density around the GRB progenitors (\citealt{Panaitescu2001}).
Also the circumburst density profile is an indicator on the nature of the
explosion (e.g., \citealt{Schulze2011}). In addition, the distribution
function of the jet-opening angles of long and short bursts should be
different from each other since an extended massive envelope collimates
the escaping relativistic outflow \citep{Zhang2004ApJ608}, while the lack of
such a medium in the case of merger events might allow for wider jet-opening
angles \citep{Aloy2005,Rezzolla2011}. Any short-burst afterglow that adds
information here is naturally of great interest.

Here we report on the results of the first  3.5 years of follow-up
observations of short-duration GRBs  using the optical/NIR seven-channel
imager GROND (\citealt{Greiner2007Msngr,Greiner2008}) mounted at the 2.2-m
ESO/MPG telescope on La Silla (Chile). GROND is in continuous operation since
mid-2007. Since then it observes every burst with a declination $\lesssim
+35^\circ$, providing a complete sample of events observed with the same
instrument at the same telescope. The capability of GROND to observe in seven
bands simultaneously, from $g'$ to $K_s$, does not only provide the
opportunity to follow the color evolution of an afterglow but also allows for
a stacking of all bands; in particular a white-light image in $g'r'i'z'$
reaches a fainter detection threshold. In addition, GROND's routine operation
in Rapid Response Mode in principle allows us to start observations within
minutes after a trigger, catching also afterglows even if they are fading
rapidly.

In this work, we summarize the detections and upper limits for 20 short burst
afterglows in $g'r'i'z'JHK_s$.  First results have already been published in
\citet[][ in the following paper I]{NicuesaGuelbenzu2011a} and
\citet[][ in the following paper II]{NicuesaGuelbenzu2012a}. Here we add
detailed information on all individual bursts. In particular, we compare the
afterglow luminosities with those of their long-burst relatives. We also
include X-ray data in order to extend this discussion to the high-energy
band. If possible, based on our optical data, we derive the spectral energy
distribution (SED) of the afterglows and give an estimate of the corresponding
jet half-opening angles.

Throughout the paper, we adopt a concordance $\Lambda$CDM cosmology
($\Omega_M=0.27$, $\Omega_{\Lambda}=0.73$, $H_0=71$~km/s/Mpc;
\citealt{Spergel2003}), and the convention that the flux density is described
as $F_\nu (t)\propto t^{-\alpha}\,\nu^{-\beta}$.  In cases where no redshift
is known for a burst, we adopt a redshift of $z$=0.5, as it is justified
based on the redshift distribution of short bursts detected 
by \swift \ by the end of 2010 (\citealt{Leibler2010}, their table 1).

\section{Target selection, observations, and data reduction}

Between July 2007 and December 2010 altogether 394 GRBs were  
localized at the arcmin or (mostly) arcsec 
scale.\footnote{http://www.mpe.mpg.de/$\sim$jcg/grbgen.html \label{foot1}}
Among them 220 events were followed up with GROND. For the present study,
from this data base we have selected all those bursts with a
duration of $T_{90} \leq 2$~s (within 1 $\sigma$) and an error circle smaller
than 3 arcmin in radius (Table~\ref{tab:coords}), giving us 20 targets.

\input{Tab1}

All optical/NIR data were analysed through standard PSF photometry using
DAOPHOT and ALLSTAR tasks of IRAF (\citealt{Tody1993}), in a similar way to
the procedure described in \citet{Thomas2008} and
\citet{Aybueke2008AIPC.1000}. PSF fitting was used to measure the magnitudes
of an optical transient. For completeness, publicly available archives were
also checked (VLT/FORS and Gemini/GMOS). 

The optical data were calibrated against the Sloan Digital Sky Survey (SDSS
DR7; \citealp{Abazajian2009}), if available. Otherwise a standard star field
was observed under photometric conditions. For the NIR bands, photometric
calibration was always performed against the 2MASS catalogue
\citep{Skrutskie2006AJ}. This procedure results in a typical absolute accuracy
of 0.04~mag in $g^\prime r^\prime i^\prime z^\prime$, 0.06~mag in  $JH$ and
0.08 mag in $K_s$. All  reported magnitudes are in the AB photometric system.
Observed magnitudes were corrected for Galactic reddening based on
\cite{Schlegel1998} and assuming a Milky Way extinction curve with a ratio of
total-to-selective extinction of $R_V=3.1$. For GROND the Vega-to-AB
conversion is  $J_{\rm AB} = J_{\rm Vega} + 0.93$~mag, $H_{\rm AB} = H_{\rm
  Vega} + 1.39$~mag, $K_{\rm AB} = K_{s,\rm Vega} + 1.80$~mag, except for
observations after an  intervention on the instrument on March 2008, for which
$K_{\rm AB} = K_{s,\rm Vega} + 1.86$~mag. Extinction corrections for the GROND
filters we have used here are: $A(g^\prime)= 1.253 \,A_V,\; A(r^\prime)= 0.799
\,A_V,\; A(i^\prime)= 0.615 \,A_V,\; A(z^\prime)= 0.454 \,A_V,\; A(J) = 0.292
\,A_V,\; A(H) = 0.184 \,A_V,\; A(K_s) = 0.136 \,A_V$. 

\section{Results}

In what follows, in several cases we combined GROND's $g^\prime r^\prime
i^\prime z^\prime$ into a \emph{white} band. This turned out to be
particularly useful when searching for a faint afterglow, for studying the
light-curve shape, and for measuring the offset of a detected afterglow from
its suspected host galaxy. Image subtraction between the first and the last
epoch, if applied, was performed using the \emph{hotpants} package.\footnote{
http://www.astro.washington.edu/users/becker/hotpants.html
\\  http://svn.pan-starrs.ifa.hawaii.edu/trac/ipp/wiki/ppSub$_-$vs$_-$Hotpants}
Errors in the astrometric accuracy of GROND are less than 0\farcs3 in right
ascension and declination. 

\subsection{GRBs with an afterglow detected by GROND \label{Sect1:OTs}}

Among the 20 events followed up by GROND, in six cases an optical afterglow
was detected by GROND. Two of these events, GRB 090426 and GRB 090510, have been
discussed in detail in paper I and II. Here we report on the four additional 
cases.

\subsubsection{GRB 081226A: Discovery of the optical afterglow 
\label{081226.txt}}

\begin{figure}[t]
\includegraphics[width=9.0cm,angle=0]{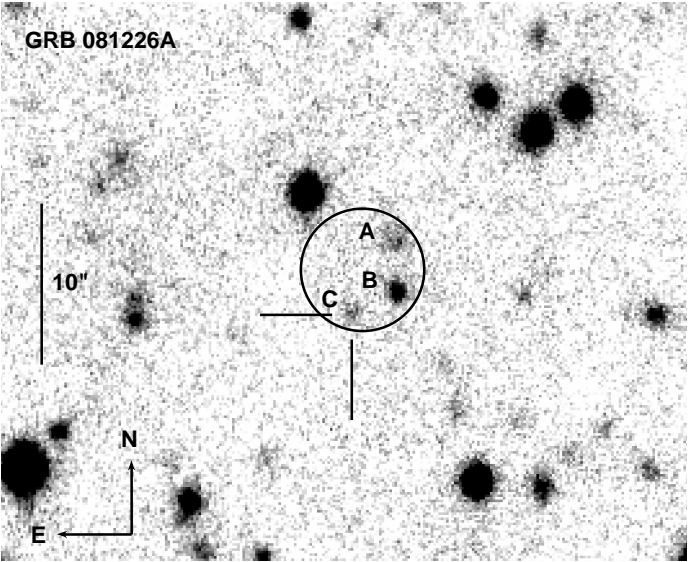}
\caption{\emph{White}-band image of the field of GRB 081226A.
Inside the 90\% c.l. XRT error circle ($r=3\farcs8$)
lie three objects (A,B,C). The position of the afterglow is indicated (C).
In order to go deep, all GROND images of the first and the second
epoch have been combined here.}
\label{fig:081226A}
\end{figure}

{\it Observations:} \ GROND started observations 10 min after the GRB trigger
and was on target for 2.5 hrs. Second-epoch observations were performed the
following night and a final epoch was obtained 1 month after the burst. Inside
the 90\% c.l. XRT error circle ($r=3\farcs8$; 
\citealt{Evans12250,Evans12273}), the white-band image shows
three objects (A-C; Fig.~\ref{fig:081226A}).

\begin{figure}[t]
\includegraphics[width=9.0cm,angle=0]{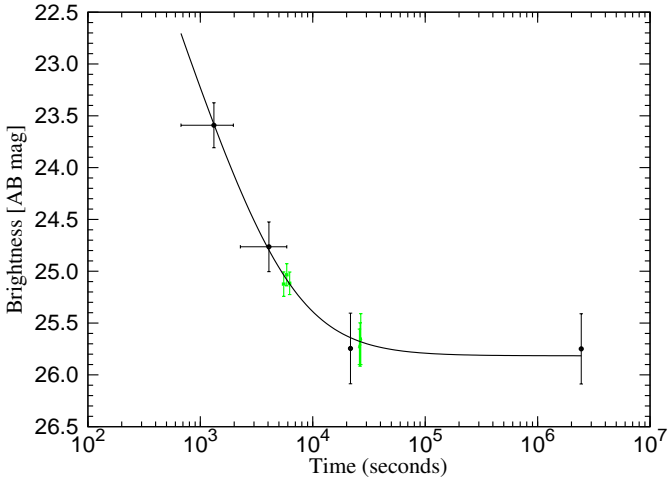}
\caption{
GROND $r^\prime$-band light curve of the afterglow of GRB 081226A fitted with
a single powerlaw plus host galaxy component. Overplotted in green color are
the Gemini-S/GMOS  $r^\prime$-band data (Table~\ref{tab:logGROND.081226A}). No
corrections for the slightly different filters have been performed.}
\label{fig:081226A_lc}
\end{figure}

{\it Afterglow light curve:} \ After performing image subtraction on the
\emph{white}-band images, the afterglow appears in the southern part of its
very faint host galaxy (object C in Fig.~\ref{fig:081226A}). It is detected in
all optical bands (Table~\ref{tab:logGROND.081226A}) and is best-sampled in
the $r^\prime$ band. Fitting the light curve with a single power-law plus host
galaxy component (Fig.~\ref{fig:081226A_lc}) gives $\alpha=1.3\pm0.2$, i.e.,
the afterglow was in the pre-jet break  evolutionary phase. The decay slope is
in agreement with the two X-ray  detections of the afterglow centered at
0.6~ks and 11.5~ks  (\citealt{Evans2010}). We re-reduced archival Gemini
$r^\prime$-band images (\citealt{Berger2008GCN8732}) and find that they fit
well into this light curve, confirming the GROND discovery.  

Due to the faintness of the afterglow, a well-defined SED, corrected for
host-galaxy light, cannot be constructed.

{\it Energy budget:} \ No redshift is known for GRB 081226A. Assuming  a
redshift of $z=0.5$ and using the  data and the numerical approach from
\cite{Butler2007}\footnote{http://astro.berkeley.edu/$^\sim$nat/swift/bat$_-$spec$_-$table.html},
we obtain an isotropic equivalent energy for this burst of $E_{\rm iso} =
2.0^{+1.7}_{-0.5}\,\times\,10^{50}$ erg. If there is a jet break in the
optical light curve then it must have occurred after about 10~ks.  Adopting an
ISM profile, for the jet half-opening angle  we have (e.g.,
\citealt{Frail2001ApJ562,Lu2011arXiv1110.4943L})
\begin{eqnarray}
\Theta_{\rm jet}&=&0.057 \, {\rm rad}\,\left(\frac{t_b}{1\ \rm
day}\right)^{3/8}\left(\frac{1+z}{2}\right)^{-3/8}\left(\frac{E_{\rm iso}}{10^{53}\ \rm
erg}\right)^{-1/8} \nonumber \\
&\times& \left(\frac{\eta_\gamma}{0.2}\right)^{1/8}\left(\frac{n}{0.1\ \rm
cm^{-3}}\right)^{1/8}\,.
\label{theta}
\end{eqnarray}
Adopting a radiative efficiency of 0.2, and scaling the results to a rather
low gas density of 0.01 cm$^{-3}$ as it might be implied for a neutron star
merger, we obtain $\Theta_{\rm jet} \gtrsim 2.6^{+0.1}_{-0.2} \,\times\,
(n/0.01)^{1/8}$ deg and a beaming-corrected energy of $E_{\rm cor} \gtrsim
2.1^{+1.3}_{-0.4}\ 10^{47} \,\times\,(n/0.01)^{2/8}$ erg. There are no X-ray
data for $t>10$ ks that could yield further evidence for a possible break in
the afterglow light curve (\citealt{Evans2010}).

{\it Host galaxy:} \ The underlying host galaxy (C)  is very faint and only
visible in the $g^\prime, r^\prime$ second-epoch images  ($g' = 25.88\pm0.24,
r' = 25.79 \pm 0.34$). The offset of the afterglow from its center is
$\lesssim$0\farcs5. For an assumed redshift of $z=0.5$ this corresponds to
$\lesssim$3 kpc. No statement can be made about the morphological type of this
galaxy.

\subsubsection{GRB 090305: Discovery of a jet break}

{\it Afterglow light curve:} \ GROND started observing the field 30 min
after the \swift/BAT trigger and was on target for 1.5 hrs.  The
fading optical afterglow \citep{Cenko2009GCN8933, Berger2009GCN8934} is
detected in all optical bands but it is not seen in the NIR
(Table~\ref{tab:ULs}). 

\begin{figure}[t]
\includegraphics[width=9.0cm,angle=0]{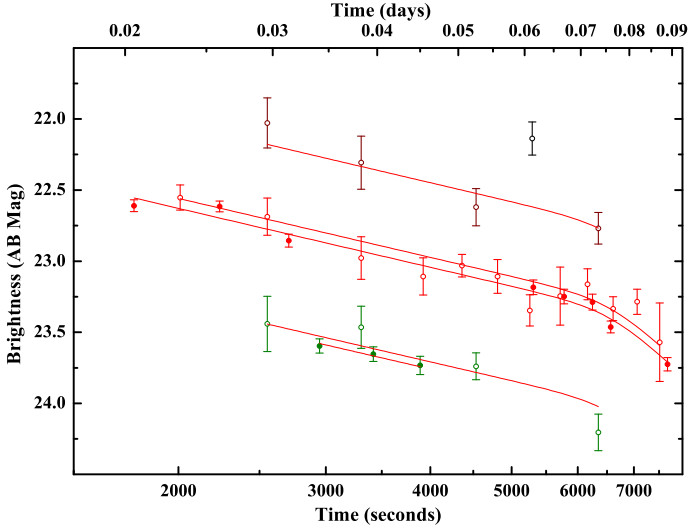}
\caption{
Gemini and GROND light curve of the optical afterglow of GRB 090305. All data
are fit simultaneously. Open circles are GROND while filled
circles are Gemini.  Color coding: green $g'$ band (shifted by $+0.5$ mag),
red $r'$ band, brown $i'$ band (shifted by $-0.5$ mag), black $z'$ band
(shifted by $-1$ mag).}
\label{fig:GROND090305lc}
\end{figure}

Gemini-S/GMOS observed from 1.5 ks to 7.5 ks after the burst in $g',r',i'$ and
discovered the afterglow (\citealt{Cenko2009GCN8933}); no detailed light curve
data have been published so far,  with $i'$-band data affected by strong
fringing. Figure~\ref{fig:GROND090305lc}  shows the result of the simultaneous
fit of all data (GROND/Gemini) using a broken power-law with the Gemini data
overplotted. The fit finds a break in the light curve at $t_b=6.6\pm0.4$ ks, a
pre-break decay slope of $\alpha_1=0.56\pm0.04$, and a post-break decay slope
of $\alpha_2 = 2.29\pm0.60$. The pre-break decay slope is rather shallow but
not unusual (e.g., \citealt{Zeh2006}). There is no X-ray light curve available
for this afterglow (\citealt{Beardmore8937}). 

\begin{figure}[t]
\includegraphics[width=9.0cm,angle=0]{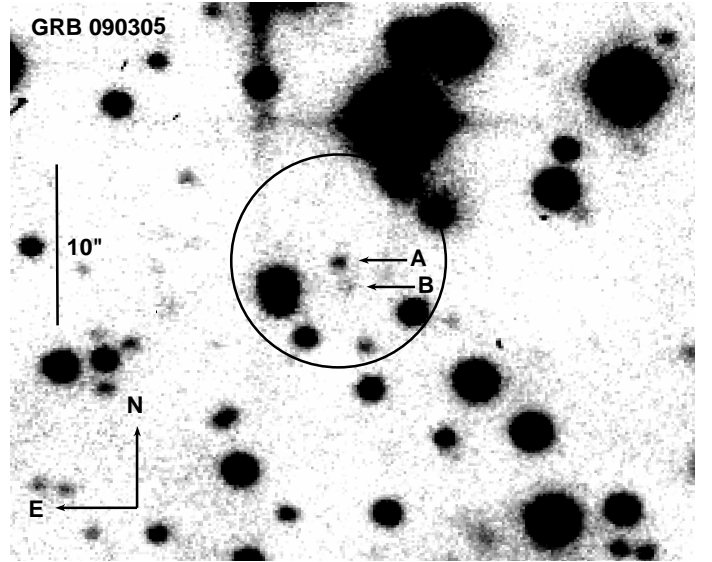}
\caption{
The field of GRB 090305: the optical afterglow (A) and the object closest to
it (B). Shown here is the $g'r'i'z'$-combined (\emph{white}-band) image taken
between 4~ks to 7~ks after the burst.  The circle is just drawn to guide the
eye; there is no independent \swift/XRT position (\citealt{Beardmore2009GCN8937}).}
\label{fig:090305host}
\end{figure}

{\it SED:} \  By fitting the Gemini $g'$ and $r'$-band data together with the
GROND $g'r'i'z'$-band data we find a spectral slope of $\beta_{\rm opt}=0.52
\pm 0.15$ ($\chi^2$/d.o.f.=0.66).  No evidence for color evolution was
found. Applying the $\alpha-\beta$  relations, there is no solution with $p>2$
for the pre-jet break phase; the light curve decay is too shallow at that time
(Table~\ref{tab:alphabeta090305}). On the other hand, the observed spectral
slope suggests that between about 2 ks and 8 ks it was $\nu_{\rm opt} <
\nu_c$, since then  $p= 2\beta+1 = 2.04 \pm 0.32$, a standard value. Possibly,
the deduced shallow $\alpha_1$ indicates that at early times the evolution of
the light curve was affected by re-brightening episodes or energy injections.
No decision can be made between a wind and an ISM model.

\begin{figure}[t]
\includegraphics[width=9.0cm,angle=0]{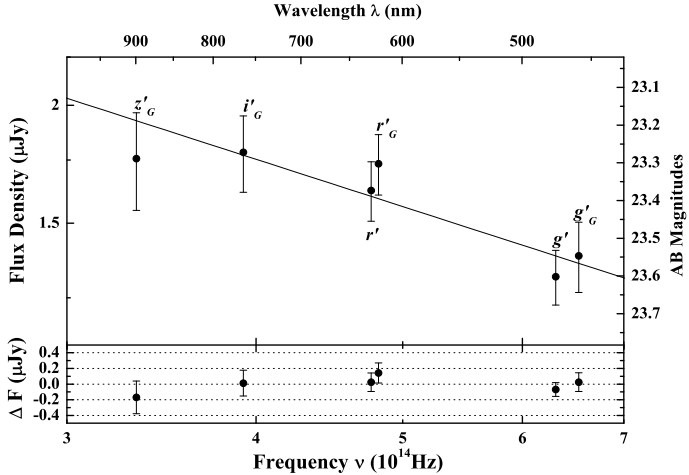}
\caption{GROND SED of the afterglow of GRB 090305 at 6 ks after the
burst, after correction for Galactic extinction. Index G stands for GROND.}
\label{fig:090305SED}
\end{figure}

\begin{table}
\caption{GRB 090305: Predicted $\beta$ based on the $\alpha-\beta$ relations
using $\alpha_1=0.56\pm0.04$ and $\alpha_2=2.29\pm0.60$.}
\begin{tabular}{cllcc}
\toprule
&&&&  \\[-2mm]   
afterglow model & $\beta(\alpha)$ & predicted $\beta$ & $p$   \\[2mm]
 \midrule
&&&& \\[-2mm]   
ISM,  iso, case 1 & $(2\alpha_1+1)/3$ & 0.71$\pm$0.06 & $1.42\pm$0.12\\ 
ISM,  iso, case 2 & $ 2\alpha_1/3   $ & 0.38$\pm$0.06 & $1.76\pm$0.12\\[1mm] 
ISM,  jet, case 1 & $\alpha_2/2$      & 1.15$\pm$0.32 & $2.30\pm$0.50\\ 
ISM,  jet, case 2 & $(\alpha_2-1)/2$  & 0.65$\pm$0.32 & $2.30\pm$0.50\\[1mm]
  \midrule \\[-3mm] 
wind, iso, case 1 & $(2\alpha_1+1)/3$ & 0.71$\pm$0.06 & $1.42\pm$0.12\\ 
wind, iso, case 2 & (2$\alpha_1-1$)/3 & 0.05$\pm$0.06 & $1.10\pm$0.12\\[1mm]
wind, jet, case 1 & $\alpha_2/2$      & 1.15$\pm$0.32 & $2.30\pm$0.50\\ 
wind, jet, case 2 & $(\alpha_2-1)/2$  & 0.65$\pm$0.32 & $2.30\pm$0.50\\[1mm]
\bottomrule       
\end{tabular}
\label{tab:alphabeta090305}
\vspace*{0.5mm}
{\bf Notes:}
Case 1 stands for $\nu > \nu_c$, case 2 for $\nu < \nu_c.$ In the former case
the power-law index of the electron distribution function 
is given by $p=2\beta$, whereas in the latter
case $p=2\beta+1$ (\citealt{Sari1999}). 
\end{table}

{\it Energy budget:} \ Assuming a redshift of $z=0.5$, and following the same
procedure as in Sect.~\ref{081226.txt}, we find  $E_{\rm iso} =
2.1^{+1.7}_{-0.7}\,\times\,10^{50}$ erg. The observed break time, if
interpreted as a jet break in an ISM medium (Eq.~\ref{theta}), leads to a jet
half-opening angle of $\Theta_{\rm jet} = 2.2^{+0.2}_{-0.1} \,\times\,
(n/0.01)^{1/8}$ deg and a beaming-corrected energy release of $E_{\rm cor} =
1.6^{+0.9}_{-0.4}\ 10^{47}\,\times\,(n/0.01)^{2/8}$ erg.

{\it Host galaxy:} \ At the position of the optical transient there is no
evidence for  an underlying host galaxy in any band, only upper limits can be
given ($g'r'i'z'JHK_s >$ 25.7, 26.0, 24.5, 24.2, 22.4, 22.0, 20.6).  The
object closest to the optical afterglow is a faint source at a distance
of 1\farcs4 (object B; see Fig.~\ref{fig:090305host}). This object is only
detected in the GROND $i^\prime$ band with a magnitude of $24.1\pm0.2$. Object
B is also detected in Gemini $r^\prime$ band data taken 10 days after the
burst at a magnitude of $26.0\pm0.1$. It was also imaged with
VLT/FORS in $R_c$ (program ID 082.D-0451; PI: A. Levan). 

Following the procedure described in \cite{Bloom2002a} and \cite{Perley2009},
the probability to find a galaxy as bright as object B within 1\farcs4
distance from the afterglow is about 7\%. Formally, this small probability
makes B a host galaxy candidate. If its observed color ($r^\prime -i^\prime$=
$2.3\pm0.2$ mag) is due to the redshifted stellar 4000~\AA \ bump, then its
redshift is around $z=0.5$.\footnote{Assuming that this is the GRB host
galaxy, this  color cannot be the Lyman break since the afterglow was
detected in the $g'$ band (\citealt{Cenko2009GCN8933}).}  For $z=0.5$
the projected distance of the afterglow from object B would then be 8.5 kpc.

\subsubsection{GRB 090927: A wind medium?}

{\it Observations:} GROND started observations  about 17 hrs after the burst
and continued for 1.5 hrs. A second-epoch observation was performed the
following night for about 1 hr. Both observing runs
were affected by bad seeing (2\farcs3). The afterglow is clearly fading in all
GROND optical bands, while it was not detected in the NIR.

\begin{figure}[t]
\includegraphics[width=9.0cm,angle=0]{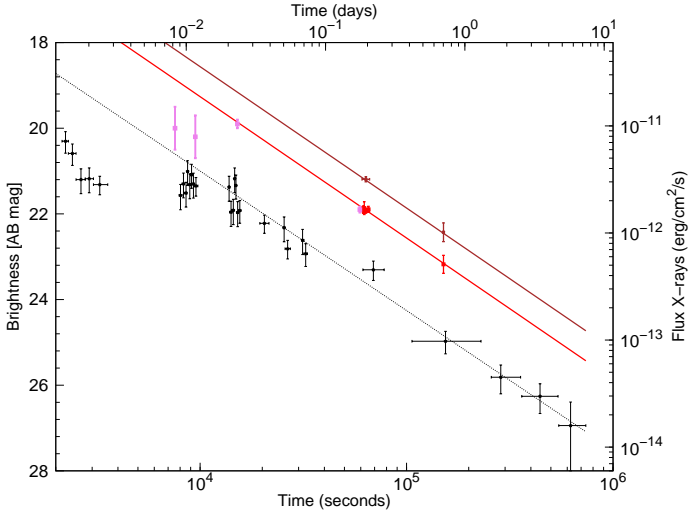}
\caption{ 
The $r',i'$-band light curve of the optical afterglow of GRB 090927 (the
$i'$-band is shifted by $-$0.6 mag;
Table~\ref{tab:logGROND.090927}). Overplotted are also $R$-band data reported
in GCN Circulars (\citealt{Klotz2009GCN9956,Cano2009GCN9960,Levan2009GCN9958};
in violet) as well as the X-ray data \citep{Evans2010}.}
\label{fig:090927lc}
\end{figure}

{\it Afterglow light curve:} \  The GROND $r^\prime$-band light curve can be
fitted with a single power-law that has a slope of
$\alpha=1.32\pm0.14\ (\chi^2$/d.o.f. = 0.39; Fig.~\ref{fig:090927lc}), which
is also in agreement with the results from the Faulkes Telescope South
(\citealt{Cano2009GCN9960}) and the VLT (\citealt{Levan2009GCN9958}). The
first two $R$-band data points from the Zadko telescope, however
(\citealt{Klotz2009GCN9956}; see appendix), lie about 1 mag below the
extrapolated fit (but also have large errors). Those data suggest that between
2 and 4 hours after the burst the optical flux was nearly constant. At the
same time the X-ray  light curve shows strong fluctuations but also seems to
be in a plateau phase. 

Assuming for the X-ray light curve a single power-law decay, for $t>20$ ks we
obtain $\alpha_{\rm X} =1.30\pm0.07$. On the other hand, the outlier at 70 ks
could also be interpreted as evidence for a break in the X-ray light curve.
However, the light curve decay after the break is then too shallow for a
post-jet  break decay slope. We thus conclude that also the X-ray afterglow is
best described by pre-jet break evolution up to the end of the XRT
observations.  A decay slope of 1.3 is in agreement with the ensemble
statistics of pre jet-break decay slopes for long-burst afterglows
\citep{Zeh2006}. 

\begin{figure}[t]
\includegraphics[width=9.0cm,angle=0]{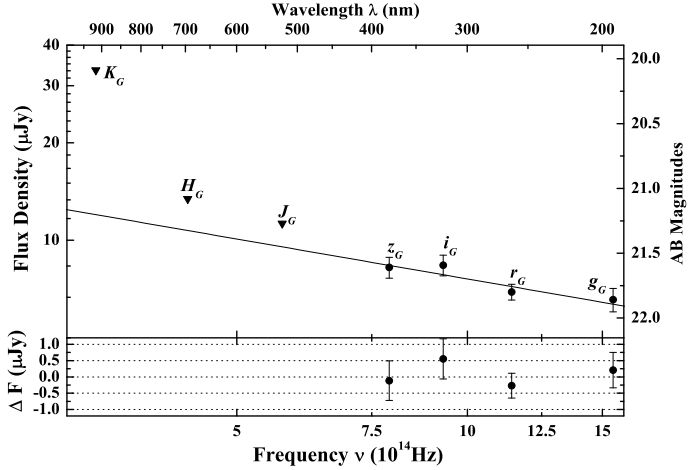}
\caption{
SED of the  afterglow of GRB 090927 at $t=$65 ks (from $g^\prime$ to $K_s$). 
It is best fit by a power law with no evidence for extinction in the host
galaxy. Note that the NIR bands are only upper limits.}
\label{fig:090927SED}
\end{figure}

{\it SED:} \ The SED of the afterglow was constructed  by combining all GROND
data taken from 64~ks to 66~ks after the trigger,  when the seeing was best
(about $2''$). It is best fit by a power law with no extinction in the host
galaxy  ($A_V^{\rm host}=0$; Fig.~\ref{fig:090927SED}). The spectral slope is
$\beta_{\rm opt}=0.41\pm0.16$. The $\alpha-\beta$-relations then imply that at
the time of the GROND observations it was $\nu_{\rm opt} < \nu_c$
(Table~\ref{tab:alphabeta090927}). The spectral slope $\beta_{\rm X}$ in the
X-ray band during this time period was $1.2\pm0.2$ (\citealt{Evans2010}),
which in combination with the spectral slope in the optical  points to
$\nu_{\rm opt} < \nu_c < \nu_{\rm X}$ and prefers a wind over an ISM
model. For the pre-jet break phase this order in frequencies implies
$\alpha_X-\alpha_{\rm opt} =\pm 0.25$ ($-$ for a wind, + for an ISM), while we
measure a difference of $-0.02\pm0.17$,  not favoring any of both models. 

\begin{figure}[t]
\includegraphics[width=6.7cm,angle=90]{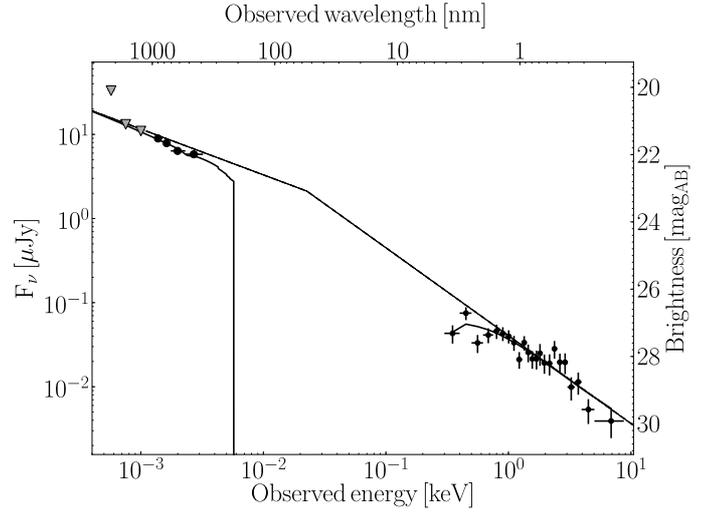}
\caption{Optical-to-X-ray SED of the  afterglow of GRB 090927 at $t=$65 ks.}
\label{fig:090927XOSED}
\end{figure}

Figure~\ref{fig:090927XOSED} shows the optical-to-X-ray SED of the afterglow
at $t$=65 ks.  Using a Galactic $N_{\rm H} = 2.9\,\times\,10^{20}$ cm$^{-2}$,
for the given  redshift ($z=1.37$; \citealt{Levan2009GCN9958})
the fit finds no evidence for host extinction (SMC dust; 
$A_V^{\rm host}=0.02\pm0.02$ mag), a spectral slope $\beta_{\rm opt}=
0.57_{-0.10}^{+0.17}$, and a break energy of 42 eV ($\chi^2$/d.o.f.=196/229
= 0.85). A fit with a single power-law is worse, confirming that  $\nu_{\rm
  opt} < \nu_c < \nu_{\rm X}$. 

{\it Energy budget:} \ Assuming a wind model, it is (\citealt{Bloom2003ApJ594})
\begin{eqnarray}
\Theta_{\rm jet}&=&0.169 \, {\rm rad}\,\left(\frac{t_b}{1\ \rm
day}\right)^{1/4}\left(\frac{1+z}{2}\right)^{-1/4}\left(\frac{E_{\rm iso}}{10^{52}\ \rm
erg}\right)^{-1/4} \nonumber \\ 
&\times& A_\star^{1/4}\ \left(\frac{\eta_\gamma}{0.2}\right)^{1/4}\,,
\label{thetaWind}
\end{eqnarray}
where $A_\star$ is the wind density parameter (\citealt{Chevalier2000ApJ536})
and, similar to Eq.~(\ref{theta}), we have introduced a radiative efficiency
$\eta_\gamma$.  For a jet-break time of $t_b>6\,\times\,10^5$~s (as implied by
the X-ray data), then for $z=1.37$ and $\eta_\gamma =0.2$, with $E_{\rm iso} =
4.5^{+3.0}_{-2.0}\,\times\,10^{51}$ erg, we find $\Theta_{\rm jet} \gtrsim
12\pm2$ deg and $E_{\rm cor} \gtrsim 1.0^{+0.3}_{-0.2} \,\times\,10^{50}$
erg. An ISM model (Eq.~\ref{theta}) gives  $\Theta_{\rm jet} =
7.0^{+0.5}_{-0.4} \,\times\, (n/0.01)^{1/8}$ deg and $E_{\rm cor} =
3.4^{+1.5}_{-1.2}\ 10^{49}\,\times\,(n/0.01)^{2/8}$ erg.

\begin{table}
\renewcommand{\tabcolsep}{3.4pt}
\caption{GRB 090927: Predicted $\beta$ based on the $\alpha-\beta$ relations
using $\alpha=1.32\pm0.14$ (for details see Table~\ref{tab:alphabeta090305}).}
\begin{center}
\begin{tabular}{cllccc}
\toprule
&&&&& \\[-2mm]   
afterglow model & $\beta(\alpha)$ & $\beta_{\rm opt}$ & $p$ & $\beta_{\rm X}$\\[2mm]
 \midrule
&&&&& \\[-2mm]   
ISM,  iso, case 1 & $(2\alpha_1+1)/3$ & 1.21$\pm$0.09 & $2.42\pm$0.18 & 1.20$\pm$0.05\\ 
ISM,  iso, case 2 & $ 2\alpha_1/3   $ & 0.88$\pm$0.09 & $2.76\pm$0.18 & 0.87$\pm$0.05\\[1mm] 
\midrule \\[-3mm] 
wind, iso, case 1 & $(2\alpha_1+1)/3$ & 1.21$\pm$0.09 & $2.42\pm$0.18 & 1.20$\pm$0.05\\ 
wind, iso, case 2 & (2$\alpha_1-1$)/3 & 0.55$\pm$0.09 & $2.10\pm$0.18 & 0.53$\pm$0.05\\[1mm]
\bottomrule       
\end{tabular}
\label{tab:alphabeta090927}
\end{center} 
\vspace*{-4mm}
Note: $p$ is given based on $\beta_{\rm opt}$. 
\end{table}

\begin{figure}[t]
\includegraphics[width=9.0cm,angle=0]{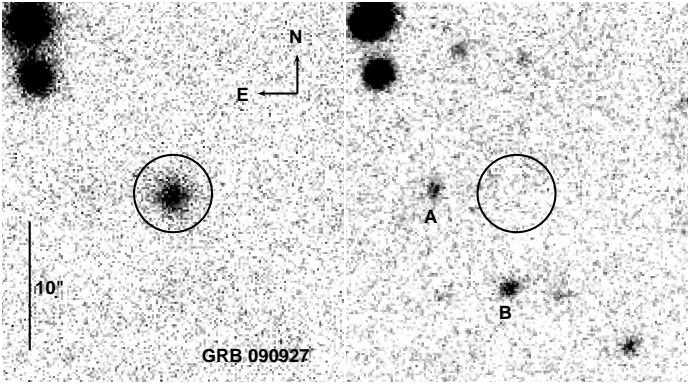}
\caption{
Finding chart of the field of GRB 090927 (GROND $g^\prime r^\prime i^\prime
z^\prime$-band combined).  \emph{Left:} First-epoch detection of the afterglow
with GROND. \emph{Right:} Deep, late-epoch observation of the field in June
2011. The circle (2\farcs5 in radius), drawn to guide the eye, is centered at
the position of the optical afterglow.  A and B label the two galaxies nearest
to the afterglow.}
\label{fig:090927host}
\end{figure}

{\it Host galaxy:} \ Observations performed two years after the trigger show
no evidence of a host galaxy at the position of the optical transient down to
deep upper limits $(g'r'i'z'JHK_s >$ 25.2, 25.2, 24.5, 24.2, 22.3, 21.6,
20.4).  The late-epoch data reveal that there are two objects (A and B) within
a radius of 10 arcsec centered at the position of the optical afterglow
(Fig.~\ref{fig:090927host}). Both objects are clearly extended. If one of them
is the host  then the projected offset of the burst was 6\farcs5 and 7\farcs5,
respectively. For a redshift of $z=1.37$ (\citealt{Levan2009GCN9958}) this
corresponds to a  projected distance of 55 kpc and 63 kpc, respectively. If
the progenitor of GRB 090927 was a collapsar, this large distance rules out
that A or B is the putative host.

\subsubsection{GRB 100117A: Determination of the afterglow decay slope}

\begin{figure}[t]
\includegraphics[width=9.0cm,angle=0]{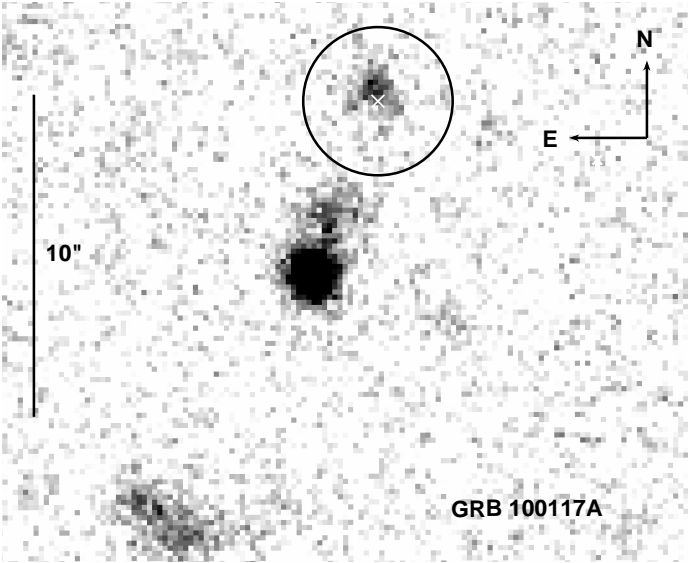}
\caption{
Combined GROND $g^\prime r^\prime i^\prime z^\prime$-band (white)  image of
the field of GRB 100117A taken half a year after the burst.  The circle is
just drawn to guide the eye, it is centered at the position of the optical
transient discovered by \cite{Fong2010} and circumscribes the GRB host
galaxy.}
\label{fig:100117Ahost}
\end{figure}

{\it Observations:} \ GROND started observing the field of GRB 100117A 3.5 hrs
after the GRB trigger and was on target for one hour
(Fig.~\ref{fig:100117Ahost}). The host galaxy flux was measured  half a year
later. 

{\it Afterglow:} \ The optical afterglow on top of its host galaxy was
discovered by \cite{Fong2010}. During the first night for the host plus
afterglow we measure a $g^\prime,r^\prime$-band magnitude of $24.37\pm 0.25,
23.72\pm 0.18$, while in the late-epoch data $g^\prime,r^\prime=25.44\pm 0.37,
24.60\pm 0.35$, resulting in a decay between both epochs of $1.07\pm0.45$ mag
and $0.88\pm0.39$ mag, respectively. 

\begin{figure}[t]
\includegraphics[width=9.0cm,angle=0]{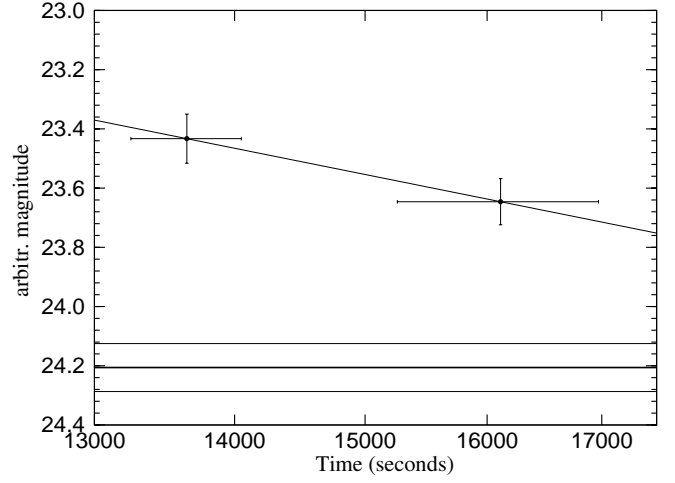}
\caption{
Combined GROND $g^\prime r^\prime i^\prime$  white-band light curve of the
decaying afterglow  of GRB 100117A, centered on galaxy
(Fig.~\ref{fig:100117Ahost}).  Also shown is the host galaxy magnitude as a
straight line, including the $1\sigma$ error (measured by GROND eight months
after the burst).  Note that the $y-$axis shows arbitrarily magnitudes.}
\label{fig:100117ALC}
\end{figure}

The second-epoch data can be used to remove the host galaxy flux from the
first epoch data.  Based on this result, we obtain an afterglow magnitude of
$r^\prime\sim 24.3$ during our first-epoch observations at a mean time of
$t=4.3$ hr. We can estimate the decay slope of the afterglow light curve by
comparing this result with the $r$-band detection of the afterglow by
\cite{Fong2010} 8.3 hrs after the burst. This gives $\alpha\sim 1.3$, assuming
no color transformation between both filters.  This result is confirmed by
combining the GROND $g'r'i'$ images into a \emph{white} band.
Figure~\ref{fig:100117ALC} shows the corresponding light curve of the
afterglow during the first night, providing a slope of $\alpha=1.2$,
indicating that during this time period the afterglow was still in its pre
jet-break phase.

{\it Energy budget:} \ \swift/XRT data do not cover the time period when GROND
and \cite{Fong2010} were observing. The last X-ray detection is at
$477^{+101}_{-57}$~s  after the trigger \citep{Evans2010}. In particular,
since the very last XRT data point at around 0.5~d is only an upper limit,
optical and XRT data cannot be compared. If the afterglow was in the pre-jet
break decay phase until at least 8.3 hr after the burst, in combination with
the observed isotropic equivalent energy of $E_{\rm iso} =
51.0^{+0.1}_{-0.1}\,\times\,10^{50}$ erg \citep{Kann2011} and a redshift of
$z=0.92$ \citep{Fong2010}, the lower limit on the jet half-opening angle is
(Eq.~\ref{theta}) $\Theta_{\rm jet} = 2.4\,\times\, (n/0.01)^{1/8}$ deg and
$E_{\rm cor} \gtrsim 4.6\  10^{48}\,\times\,(n/0.01)^{2/8}$ erg. 

{\it Host galaxy:} \ Our data do not allow us to measure
the  offset of the afterglow from its host galaxy center; \cite{Fong2010},
using their Gemini-N/GMOS observations, obtained 60$\pm$40 mas,
corresponding to $0.5\pm0.3$ kpc.

\subsection{GRBs with no afterglow detected by GROND \label{Sect2:NOTs}}

The results for those 14 out of 20 GRBs where GROND could not detect the
afterglow are summarized in Table~\ref{tab:ULs}.  In most cases  we were on
target within some hours after the burst. In all cases deep upper limits can
be provided, in particular in the NIR, where we reach up to $J$=22.7,
$H$=22.0, and $K_s$=21.2. The individual observations by GROND are described
in detail in the appendix. However, of particular interest are two events
(GRB 080919, 100702A), where observations started within less than 10 min
after the trigger.

\input{Tab2}

\subsubsection{GRB 080919 \label{080919.txt} }

GROND started observing the field  8 min after the burst. Due to a delay
in secure XRT coordinates (\citealt{Preger2008GCN8270}), during the  first 30
min  only the NIR images cover the afterglow position. Deep second-epoch
observations  were performed with GROND three years after the burst.  Image
subtraction was performed between second and first-epoch data in all bands but
no afterglow was found.  Probably the main reason for this non-detection is
the presence of a bright star inside  the error circle which makes it
difficult to detect any faint transient in spite of the small XRT error circle
(90\% c.l. radius $r=1\farcs6$;  \citealt{Evans12273}).  Therefore, we note
that the upper limits we provide in Table~\ref{tab:ULs} refer to isolated
objects in the field while the more reliable upper limits for the afterglow
can be substantially less deep than reported there.

\begin{figure}[t]
\includegraphics[width=8.4cm,angle=0]{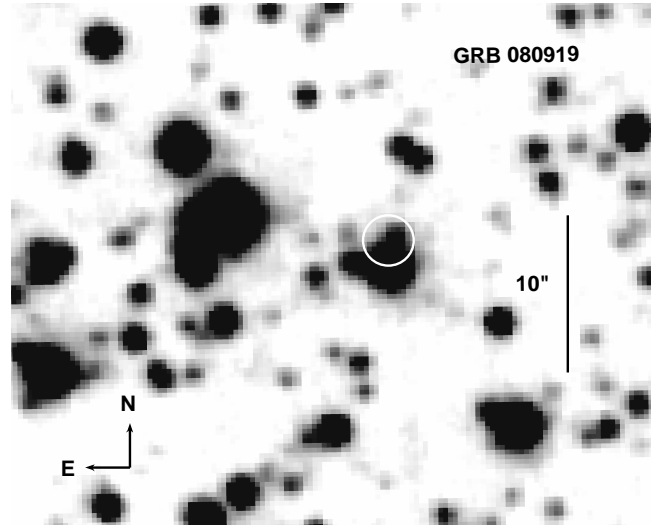}
\caption{
The field of GRB 080919. The XRT error circle (radius $r=1\farcs6$) lies 
close to a relatively bright foreground star.}
\label{fig:080919psfsub}
\end{figure}

\subsubsection{GRB 100702A}

\begin{figure}[t]
\includegraphics[width=9.0cm,angle=0]{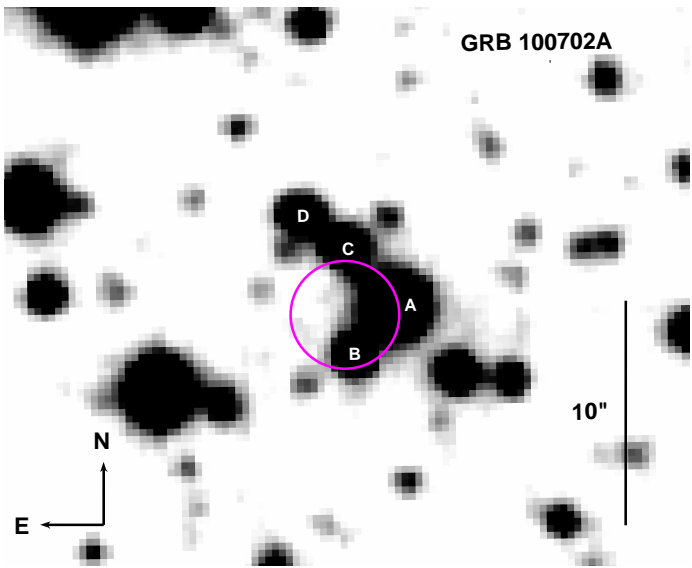}
\caption{Finding chart of the field of GRB 100702A in the GROND $J$ band. 
Shown also is the 90\% c.l. XRT error circle ($r=2\farcs4$; 
\citealt{Siegel2010GCN10916}).}
\label{fig:100702chart}
\end{figure}

GROND started to observe the field 2.5 min after the burst. Inside the 90\%
c.l. XRT error circle  ($r$=2\farcs4; \citealt{Siegel2010GCN10916}) the GROND
data reveal two bright objects (A, B) within the XRT error circle  and two
others (C, D) close by (Fig.~\ref{fig:100702chart}; see also
\citealt{Malesani2010GCN10918}).  Objects A and B look have a point-like PSF
and might be stars, while C and D might be galaxies.

Image subtraction and PSF photometry in each  band was performed for all
objects but no evidence for variability was found, neither in the optical nor
in the NIR bands; only upper limits can be provided for any afterglow
(Table~\ref{tab:ULs}). Similarly to GRB 080919, the upper limits refer to
isolated objects in the field.

\section{Discussion} 

Including our discovery of the afterglow and host galaxy of GRB 081226A, nine
out of 20 short-bursts in our sample have a discovered optical transient,
while six have only a \swift/XRT and four have only a BAT/IBIS localization
with no optical afterglow. Among the 9 bursts with detected optical transient
six events have a redshift reported in the literature. An additional redshift
information comes from the identification of the host galaxies in the case of
GRBs 100206A \citep{Cenko10389, Perley2011}, 100628A \citep{Cenko10946}, and
101219A \citep{Chornock11518}. These redshifts range from $z=0.10$ (GRB
100628A) to $z=2.61$ (GRB 090426). Four of the 9 bursts have a redshift of
smaller than 0.5, a high percentage compared to the long-burst population; for
more redshifts of short-bursts see the compilations by \cite{Berger2009ApJ690}
and \cite{Kann2011}.

The best-sampled light curves are those of GRB 090426 (paper I) and GRB 090510
(paper II) followed by (ordered by sampling quality) GRBs 090305, 081226A,
090927, and 100117A. Only the afterglow of GRB 090426 has NIR detections. In
three cases we find a clear break in the light curve, partly in combination
with data obtained at other facilities.  Two of these events (GRBs 090426,
090510) were imaged by GROND in the post-break decay phase only and for GRB
090305 the data included also the pre-break phase. In principle, the three
breaks might be interpreted as jet breaks but for GRB 090510 the \swift/UVOT
data suggest a different explanation, namely the passage of the injection
frequency across the GROND bands (for details see \citealt{Kumar2010},
\citealt{DePasquale2010} and paper II). For the other three cases the light
curves can be fitted with a single power law and, based on the deduced decay
slope, observations were performed during the pre-jet break evolutionary
phase.  The light curve decay slopes as well as the spectral slopes are not
different from what is known for the long-burst sample
(Table~\ref{tab:summary}).

\subsection{Optical luminosities}

In the last years, evidence has been mounting that the classical $T_{90}$
division between short and long GRBs is not transferable to a more physically
inspired division between progenitor models. It seems that merging compact
objects may result in high-energy emission on timescales far exceeding
$T_{90}=2$ s, whereas conversely collapsar-triggered GRBs can be luminous
short spikes with $T_{90,rest}<2$ s. This led \cite{Zhang2007} to propose,
analogous to the designations of supernovae, that GRBs come in two types: Type
I GRBs stem from the coalescence of massive compact objects, whereas Type II
GRBs are associated with the core-collapse of massive stars. \cite{Zhang2009}
studied the observational signatures of the two classes and devised a scheme
to classify GRBs. \cite{Kann2011} studied a large sample of Type I
candidate GRBs, adding the optical afterglow luminosity at late times as an
additional criterion to discern the two classes, with Type I GRB afterglows
being much less luminous than those of Type II GRBs.

\begin{figure}[t]
\includegraphics[width=8.9cm,angle=0]{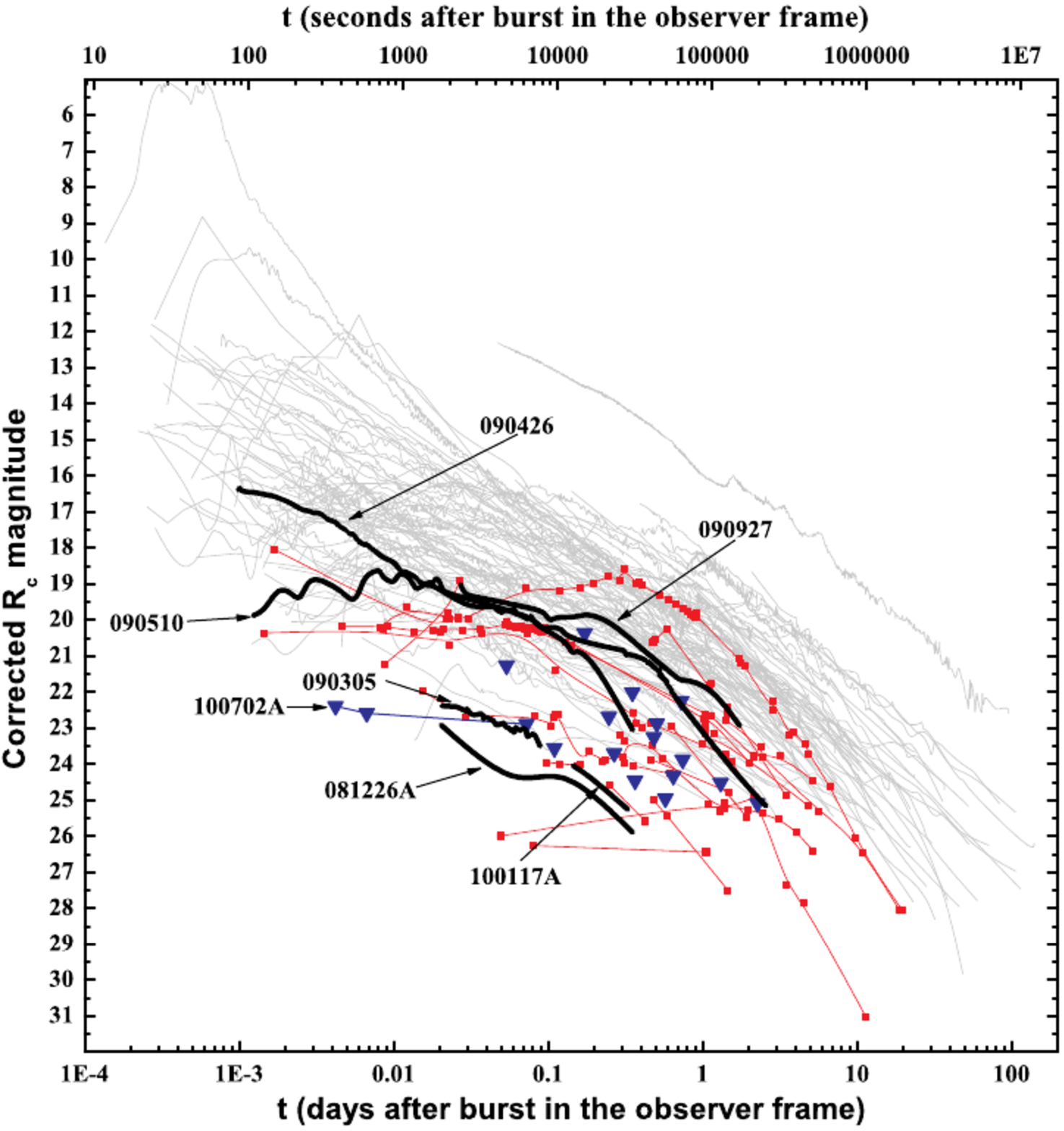}
\caption{
Light curves of long and short GRB afterglows. These light curves have been
corrected individually for Galactic foreground extinction following
\cite{Schlegel1998}, and, if possible, host galaxy contribution. The thin gray
lines are the long GRB sample of \cite{Kann2010}. The red squares connected by
splines represent the afterglow detections reported by \cite{Kann2011}. The
short GRB afterglows detected by GROND and presented in paper I and II as well
as this work are given as labeled thick black lines (they may include
additional data beyond the GROND detections). Upper limits presented in this
work (Table~\ref{tab:ULs}) are given as blue triangles. GRB 100702A is
highlighted also  because of its very early upper limits.  The last data point
for GRB 100117A is from \cite{Fong2010}, the others as well as the data for
GRBs 090305 and 081226A are presented in this paper. Early data for GRB 090927
are taken from \cite{Klotz2009GCN9956}, \cite{Levan2009GCN9958},
\cite{Cano2009GCN9960} as well as \cite{Kuin2009GCN9954}.}
\label{fig:AlexLCobs}
\end{figure}

\begin{figure*}[t]
\includegraphics[width=18.5cm,angle=0]{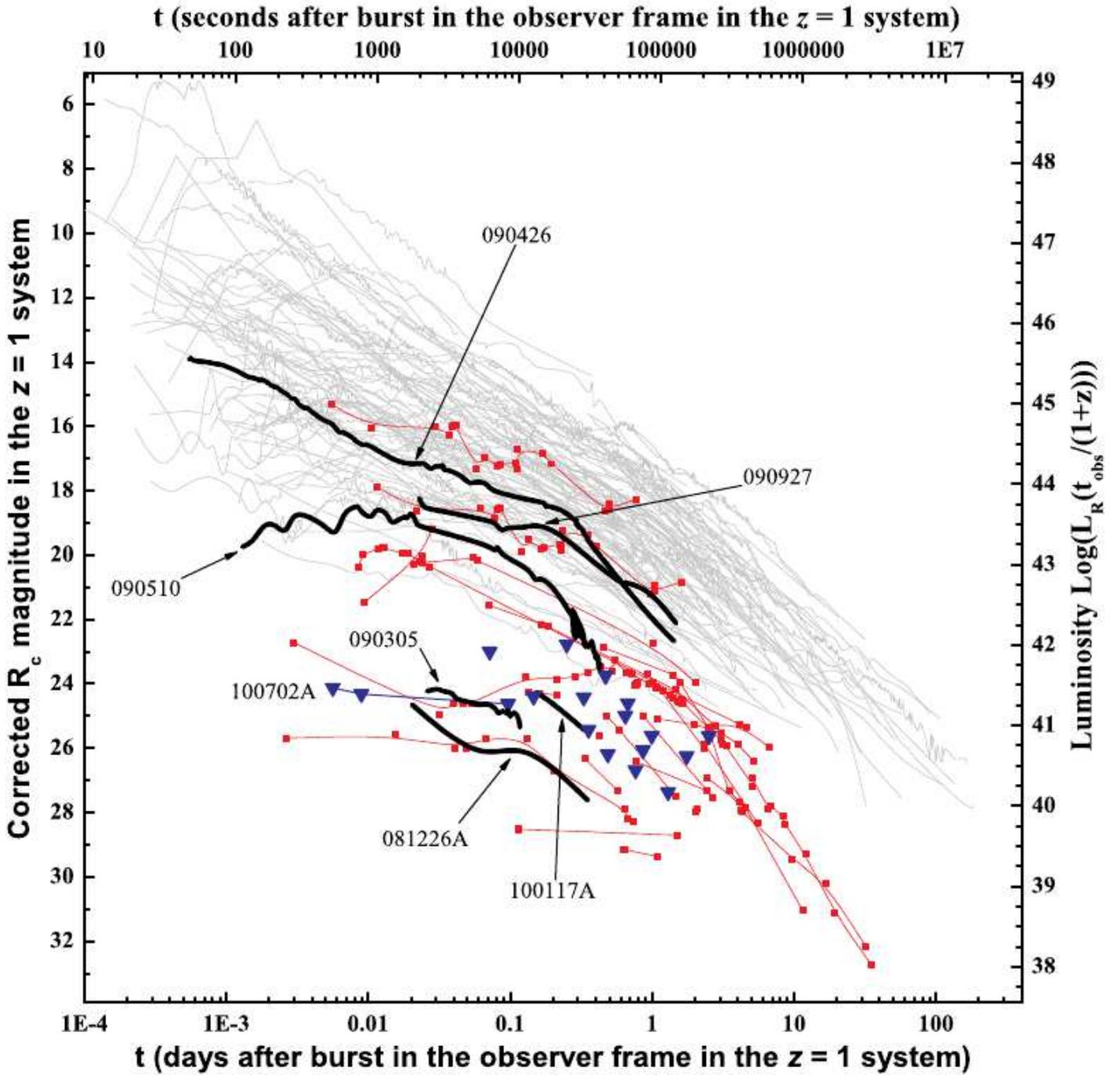}
\caption{
The light curves of the six GROND-detected short GRB afterglows as well as the
upper limits in the redshift $z=1$ frame. The labeling is identical to Figure
\ref{fig:AlexLCobs}. GRBs 090426 and 090927 are likely Type II. 
The luminosity is in units of erg/s. See text for further details.}
\label{fig:AlexLC}
\end{figure*}

So far, in this work, we have discussed the sample based on the classic
$T_{90}$ division. What can the optical luminosity of the afterglows (or upper
limits thereon) tell us about the likely progenitor systems?
Figure~\ref{fig:AlexLCobs} is a continuation of the plots shown in
\cite{Kann2006, Kann2010, Kann2011}. Against the background of Type II GRB
afterglow light curves (thin gray lines), we show the Type I GRB afterglow
detections as presented in (\citealt{Kann2011}; red squares connected by
splines, upper limits have been omitted for clarity) as well as the detected
afterglows (thick black splines) and upper limits (downward-pointing blue
triangles) derived by GROND in this work as well as in paper I and II.

Already in this plot it is visible that the short GRB afterglows are less
bright than the mean brightness of the long GRB afterglows, with half of them
(GRBs 090305, 10017A and 081226A) being as faint or fainter than the faintest
so-far detected long GRB afterglows. A true comparison needs to account for
the redshift and intrinsic extinction, though.

Figure~\ref{fig:AlexLC} shows the light curves of the six short GRBs detected
with GROND in the $z=1$ reference frame, having been corrected for both
distance and intrinsic reddening in the GRB host galaxy, if possible
\citep{Kann2006,Nardini2006}. A redshift of $z=0.5$ and zero host extinction
was assumed for all cases where these values are not known. Of the six
afterglows, that of GRB 090426 is now seen to be the most luminous, followed
by the ones of GRBs 090927 and 090510. Several arguments have already been put
forward that the origin of 090426 was a collapsar event (see paper I and
references therein). Between about 0.01 and 0.1~d after the burst (measured in
the GRB host frame), its magnitude (for the fixed distance a measure of the
luminosity) was about 2 mag brighter than the magnitude of the optical
afterglow of the other two events. The optical afterglow of GRB 090510, if due
to a merger event, must be characterized as very luminous between $\sim0.005$
and 0.1~d after the burst. Because of its emission in the 10-100 GeV band and
its outstandingly small jet half-opening angle of $\Theta_{\rm jet}\lesssim
1^\circ$ (\citealt{DePasquale2010,He2011,Kumar2010}, paper II;  if correctly
interpreted in this way), it was special in several other respects, too. The
optical afterglow of GRB 090927 reached the luminosity of the afterglow of GRB
090426 at about 1~d after the burst, but its further evolution is
unfortunately unknown. This moderately high optical luminosity along with
significant lag and other spectral characteristics
\citep{Stamatikos2009GCN9955} and a redshift beyond what is seen for Type I
GRBs \citep{Levan2009GCN9958} argue that GRB 090927 is also likely to be a
Type II GRB. All other afterglows with GROND detections or GROND upper limits
fall well within the Type I GRB sample.

\input{Tabgrbs.tex}

Between about 0.01 and 0.1~d (host frame time) the three optical afterglows
mentioned above (which have a measured redshift) were about $7 \pm 1$ mag
brighter than the afterglows of GRBs 081226A, 090305, and 100117A (among which
only the latter has a secure redshift).\footnote{if  the redshift of the
former two bursts is not 0.5, as assumed here, but  somewhere in the range
between 0.2 and 1.0, then this magnitude difference changes by about $\pm2$
mag} For GRB 090510, the situation changes after about 0.1~d, when the early
break and following steep decay (paper II) lead it to become much fainter than
the Type II GRB afterglows \citep[see also][]{Kann2011}. From the perspective
of optical luminosities, we therefore find additional evidence for a collapsar
origin of GRB 090927, despite its short duration, whereas there is no evidence
indicating that GRBs 090305 and 081226A are not members of the classical
short/Type I GRB population. We note in passing, though, that
\cite{Panaitescu2011} also discussed a collapsar origin for GRB 090510.

\subsection{Jet half-opening angles}

Observations of jet breaks in short-burst afterglow light curves are rather
sparse, in the optical as well as in the X-ray band. In the optical band, the
best-sampled cases are GRBs 090426 and 090510, but the former burst is
suspected to be due to a collapsar explosion rather than due to a merger event
(e.g., \citealt{Thoene2011MNRAS}),  while the latter stands apart even from
the long-burst sample due to its very small jet half-opening angle
(\citealt{He2011}). The third member of this group is GRB 050709 with an
estimated $\Theta_{\rm jet}\sim 14$ deg \citep{Fox2005Natur437}, which is
based on a very sparsely sampled light curve, however. 

In the X-ray band the observational situation is not much better. The best
case might again be GRB 090510 \citep{DePasquale2010}, followed by GRBs
050724, 051221A, 061201, and 111020A. Unfortunately, the first burst (GRB
050724) allows only for an estimate of a lower limit on $\Theta_{\rm jet}$
($\gtrsim 25$ deg; \citealt{Grupe2006ApJ653,Malesani2007}), while GRB 051221A
relies on a rather well-sampled light curve (leading to $\Theta_{\rm jet}
\sim$4--8 deg; \citealt{Burrows2006ApJ653,Soderberg2006}). The X-ray light
curve of GRB 061201 is well-sampled, too \citep{Stratta2007}; again the
observed break time is quite early ($\sim$40~min; $\Theta_{\rm jet}$=1--2
deg). Recently, \cite{Fong2012arXiv1204.5475F} reported on the X-ray light
curve of the short burst 111020A, which showed a break at 2~d, leading to an
estimated $\Theta_{\rm jet} = 3-8$ deg for an assumed $z$=0.5-1.5 and
$n\sim$0.01 cm$^{-3}$. 

\begin{figure}[t]
\includegraphics[width=9.0cm,angle=0]{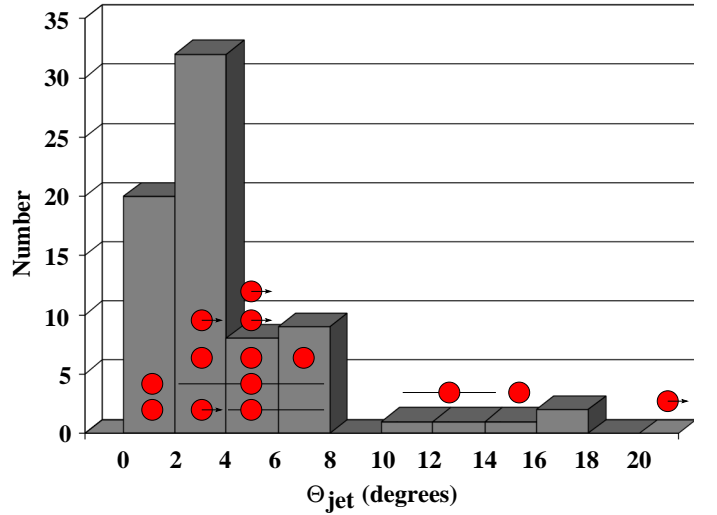}
\caption{
The observed distribution of jet half-opening angles of 74 long bursts (based on
the compilation in \citealt{Lu2011arXiv1110.4943L}) compared to the
short-burst sample. Since the latter has much  less data, we do not plot a
histogram but only points.  An arrow indicates a lower limit on $\Theta_{\rm
jet}$.  The Type I events GRB 051221A, 060614, and 070714B listed in
\cite{Lu2011arXiv1110.4943L} have not been used for the plot of the long-burst
data.}
\label{fig:Thetas}
\end{figure}

Figure~\ref{fig:Thetas} shows the observed distribution of jet half-opening
angles of long-bursts based on the compilation of \cite{Lu2011arXiv1110.4943L}
compared to the short-burst sample (a similar plot is recently shown by 
\cite{Fong2012arXiv1204.5475F}. The latter contains the results summarized
in Table~\ref{tab:summary}, supplemented by  GRBs 061006 ($\Theta_{\rm
jet}\sim 5$ deg),  070714B ($\Theta_{\rm jet}\gtrsim 4$ deg), and  071227
($\Theta_{\rm jet}\gtrsim 4$ deg)  taken from the compilation of
\cite{Fan2011} but using $\eta_\gamma=0.2$ instead of 1.0 (i.e., multiplying
their numbers by 0.8;  Eq.~\ref{theta}). At a first view, this figure shows
tentative evidence that short bursts have wider jet-opening angles than long
bursts. Some caution is necessary, however. First at all,  when
calculating the jet half-opening angles, \cite{Lu2011arXiv1110.4943L} assumed
$n=0.1$ cm$^{-3}$ and $\eta_\gamma=0.2$ throughout. Even though $\Theta_{\rm
  jet}$ is only modestly sensitive to changes in both parameters (see
Eq.~\ref{theta}), gas densities derived for bursts based on
multi-wavelength data show a spread from burst to burst by several orders of
magnitude (e.g., \citealt{Panaitescu2001ApJ554}). Second, error bars in
$\Theta_{\rm jet}$ are not taken into account in the histogram. 
Similarly, our standard assumption of $n=0.01$
cm$^{-3}$ for short  bursts is a simplification, too. Possibly for individual
bursts it can be wrong by a factor of up to 100 in both directions. Finally,
our plot contains  only long bursts with measured jet break times. A more
detailed study should also contain those long bursts for which only a lower
limit on  $\Theta_{\rm jet}$ can be given (e.g., \citealt{Grupe2007ApJ662}).

\subsection{X-ray afterglows}

We selected from the \swift \ Burst Analyser \citep{Evans2010} all
bursts with detected X-ray afterglow and measured redshift that were
detected between January 2005 and August 2011. We then shifted all
light curves to their rest frames following \citet{Greiner2009ApJ693}.
If no redshift information is available for a short-burst in our sample
(Table~\ref{tab:coords}), we assumed a redshift of $z=0.5$.

\begin{figure*}[t]
\includegraphics[width=18.0cm,angle=-90]{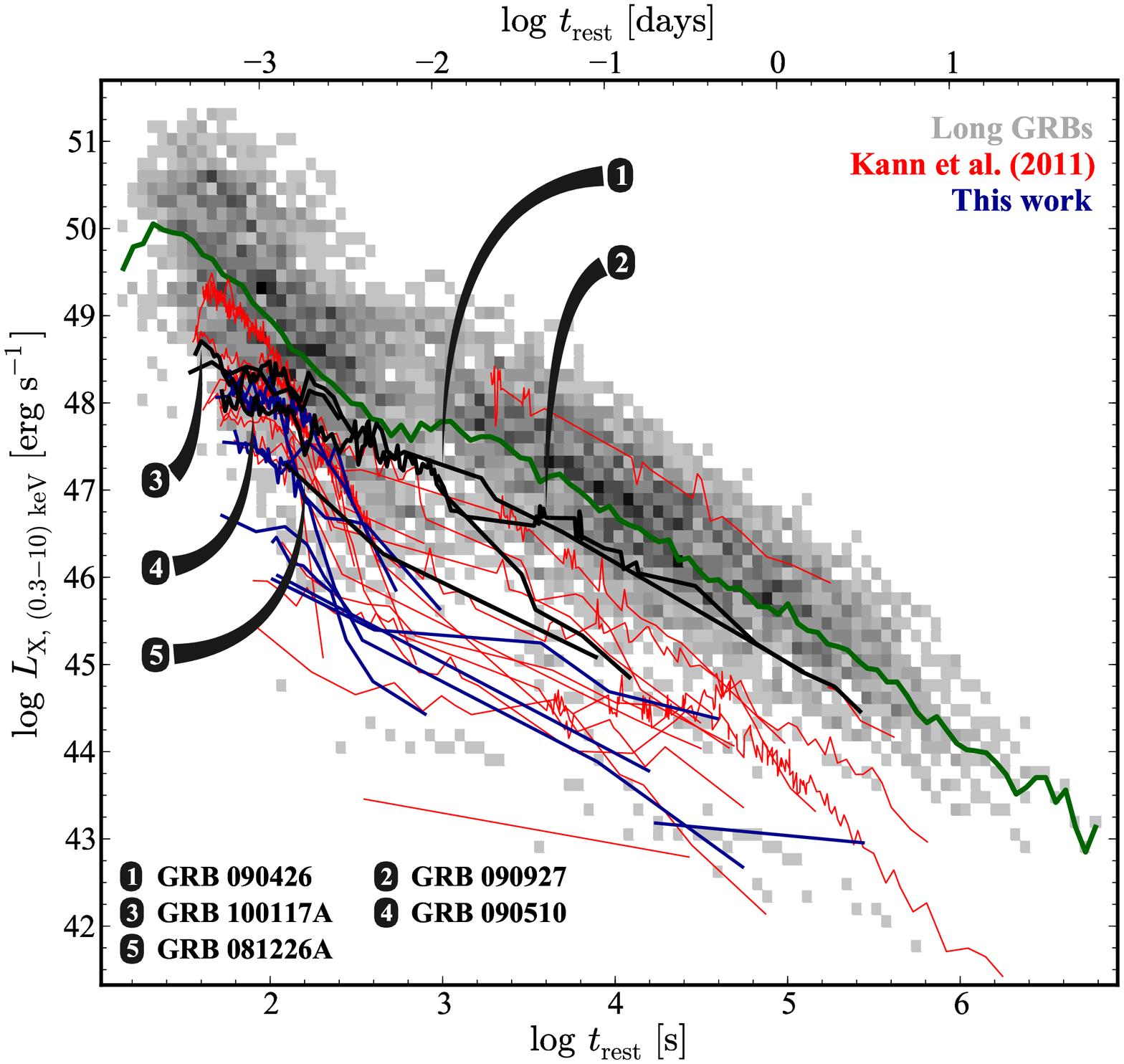}
\caption{
Shown here is the luminosity evolution of the X-ray afterglows of the
short-bursts in our sample. Highlighted are the bursts with optical afterglows
(Figs.~\ref{fig:AlexLCobs},\ref{fig:AlexLC}). Black lines represent the
afterglows of the likely  Type II events GRB 090426 and GRB 00927, green the
afterglows of the Type I events GRBs 081226A, 090510, and 100117A.  No X-ray
afterglow light curve  was reported for GRB 090305.  Overplotted is also the
short-burst sample compiled by \citet{Kann2011} (red color) as well as the
X-ray afterglows of the long-burst sample with known redshift (gray). All
short-burst afterglows are less luminous than the mean of long-burst-afterglow
luminosities (orange line), however there is a continuous overlap between
short and long GRB afterglows.}
\label{fig:XLCs}
\end{figure*}

Figure~\ref{fig:XLCs} displays the resulting luminosity evolution of
those 14 bursts in our sample for which an X-ray afterglow light curve
can be constructed, i.e., the X-ray afterglow is detected during at least
two epochs. This excludes GRBs 071112B, 081226B, 090305, 091117A, and
101129A from the plot, which have no afterglow detection at all, and it
also excludes GRB 100206A that is only detected once. The figure also
shows the luminosity evolution of 191 long GRBs with measured redshift.
In addition, we overplot the short-burst sample compiled by \citet{Kann2011},
consisting of an additional group of 19 events that are not included
in our short-burst sample.

Figure~\ref{fig:XLCs} demonstrates that the X-ray afterglows of short-bursts
represent the low end of the luminosity distribution of X-ray afterglows. They
are on average a factor of $\sim100$ less luminous than those of  long-bursts,
similar to what is seen for optical afterglows \citep[Fig.~\ref{fig:AlexLC};
see also][]{Gehrels2008, Nysewander2009, Kann2011}. However, with the single
exception of GRB 050509B, short-bursts do not represent the least-luminous
X-ray afterglows known. There is a continuous overlap between both
populations; for certain time intervals several long-burst afterglows are even
less luminous than the population of short-burst afterglows.

There is a remarkable concentration of short-burst afterglows in a relatively
narrow luminosity band around $L_{X,\,(0.3-10)\, \rm keV]}\simeq
10^{48}\,\rm{erg/s}$ at $t\sim100\,\rm s$ in the rest frame. Even after
removing bursts with assumed redshifts, the concentration is still present,
indicating that this is a genuine feature and is not an artifact caused by
bursts with assumed redshifts. After that time the luminosities of most
short-burst afterglows drop notably and their luminosity distribution broadens
by an additional factor of $\sim10$ to a final range of $\sim100$, which holds
up to at least t=1~d. At even later times most short-bursts are not 
detected anymore. Outstanding here is the X-ray afterglow of GRB 060614, which
was detected until $t=2\,\times\,10^6$~s (rest-frame), while in our sample only
three events (GRBs 090426, 090927, and 100628A) have been detected beyond
$t=10^5$~s. We caution that the former two are possibly Type II GRBs, i.e.
originating from the gravitational collapse of a massive star.

In our sample, the X-ray afterglows of the short-bursts GRBs 071227
($z=0.383$; \citealt{DAvanzo7152})  and 080905A ($z=0.122$;
\citealt{Rowlinson2010a}) have the lowest luminosities, while GRB 090927
($z=1.37$; \citealt{Levan2009GCN9958}) and 090426 ($z=2.609$;
\citealt{Levesque9264}) are the most luminous short-bursts in our sample,
again we stress that the latter two are likely Type II GRBs. Adding the data
set discussed in \citet{Kann2011}, then the X-ray afterglow of GRB 050509B
represents the low end of the luminosity distribution between $\sim0.3$~ks and
$\sim30$~ks, followed by GRBs 061201, 060505, and 060614\footnote{which is
likely a Type I GRB despite its long duration, \citealt{Zhang2009},
\citealt{Kann2011}} at later times. On the other hand, the most luminous
short-burst afterglows are that of GRBs 080503 and 051210 which reach $\log
\left(L_{X,\,(0.3-10)\,\rm keV]} [\rm{erg/s}] \right)) \simeq 49.25$ during
the peak of their emission at $\sim100$ s. Only the X-ray afterglow of GRB
060121 is more luminous at later times, assuming $z=4.6$
(\citealt{deUgartePostigo2006ApJ648}; but this GRB is possibly also Type II
GRB, \citealt{Kann2011}). 

\section{Summary}

We have reported on the results of 3.5 yrs follow-up observations of
short-duration GRBs (defined by $T_{90}<2$ s) using the multi-channel imager
GROND mounted at the 2.2-m telescope on La Silla. GROND is especially designed
to perform rapid follow-up observations of afterglows, which is particularly
useful for short-duration GRBs because of their on average very faint optical
afterglows (\citealt{Nysewander2009,Kann2010,Kann2011}). To our knowledge,
what we have presented here is one of the most comprehensive data sets  on
short-burst follow-up observations published so far, although most of them
provide only upper limits.

Among the twenty events followed-up by GROND, in six cases GROND could image
the fading optical afterglow. Five of them had already been known in the
literature (GRBs 090305, 090426, 090510, 090927, 100117A), and the GROND
follow-up observations of GRBs 090426 and 090510 were already represented in
paper I and II. The new discovery reported here is the optical afterglow of
GRB 081226A. It was imaged by GROND superimposed on its faint host galaxy
($r'\sim 25.8$) and faded away already within 10 ks after the burst. GRB
081226A also belongs to those three cases in our sample where GROND was on
target within 10 min after the trigger. The other two events (GRBs 080919 and
100702A), even though with very small X-ray error circles, were unfortunately
located in fields crowded by stars, preventing the discovery of the
optical/NIR afterglow in any band.

Three of the six optical afterglow light curves (GRBs 090305, 090426, 090510)
show a break that can be interpreted as a jet break. The other three
afterglows  (GRBs 081226A, 090927, 100117A) show a decay slope in agreement
with a pre-jet break evolution, allowing us to set at least lower constraints
on their corresponding jet half-opening angle, $\Theta_{\rm jet}$.  When
comparing these results with the long-burst population, we find tentative
evidence for wider jet-opening angles of short bursts  compared to their
long-duration relatives. However,  it might need another 20, or so,
short-burst afterglow light curves with well detected jet breaks before
observations can seriously start to constrain theoretical models. Moreover, it
should be kept in mind that some long-duration GRBs have relatively large
jet-opening angles, too (e.g.,
\citealt{Grupe2007ApJ662,Racusin2009,Liang2008}); a clear separation  between
long and short bursts with respect to their $\Theta_{\rm jet}$ values does
obviously not exist.

The separation between merger and collapsar events becomes  more evident when
the luminosities of their optical and X-ray afterglows are compared.  If the
Type I/II classification scheme is used, GRBs 090426 and 090927 have a
collapsar origin (\citealt{Kann2011}), and in fact their afterglow
luminosities in the optical band lie in the region occupied by the main body
of the long-burst/collapsar population (Fig.~\ref{fig:AlexLC}). The optical
luminosities of the afterglows of the Type I GRBs 081226A, 090305, and 100117A
are substantially smaller and stand apart from the parameter space occupied by
the long-burst sample. On the other hand, the optical afterglow of GRB 090510,
which was special due to its very high-energy emission (see appendix), seems
to be an intermediate case.

Seven years after the first precise localizations of short-duration GRBs by
\swift, the discovery of their optical afterglows remains an observational
challenge. Even though the list of well-localized short-bursts is not that
small anymore (\citealt{Nysewander2009,Kann2011}; for a continuous
update see footnote \ref{foot1}), the number of well-observed light curves
of short-burst afterglows is rather small. Progress in this respect might be
strongly linked to the availability of GRB-dedicated instruments on at least
medium-class optical telescopes. GROND is one of them.

\begin{acknowledgements}

A.N.G., D.A.K., A.R.,  and S.K. acknowledge support by grant DFG Kl
766/16-1. A.N.G., A.R., D.A.K., and A.U. are grateful for travel funding
support through the MPE.  A.R. acknowledges additional support by the 
Jenaer Graduierten\-akademie. S.S. acknowledges support by a Grant of
Excellence from the Iceland Research.  T.K. acknowledges funding  by the DFG
cluster of excellence 'Origin and Structure of the Universe', F.O.E. funding
of his Ph.D. through the DAAD, M.N. support by DFG grant SA 2001/2-1 and
P.S. by DFG grant SA 2001/1-1. Part of the funding for GROND (both hardware
and personnel) was generously granted by the Leibniz-Prize to G. Hasinger (DFG
grant HA 1850/28-1). This work made use of data supplied by the UK Swift
science data center at the University of Leicester. 
\end{acknowledgements}


\bibliographystyle{aa}
\bibliography{mypaper}

\appendix

\section{Data tables}

\begin{table}[!h]
\caption[]{Log of the GROND observations of the afterglow (plus host) of 
GRB 081226A (Fig.~\ref{fig:081226A_lc}). These results supercede the data given in \cite{Afonso2008GCN8731}.}
\renewcommand{\tabcolsep}{1.5pt}
\begin{center}
\begin{tabular}{llllllll}
\toprule
 Time (s) &$g^{\prime}$ &$r^{\prime}$ & $i^{\prime}$&$z^{\prime}$  &  $J$  &   $H$    &        $K_s$         \\
\midrule
  1320   & $>$ 24.1   & 23.59(22) & $>$ 23.1   & 22.86(35)    &  $>$20.9    &   $>$ 20.3   &   $>$19.6 \\
  4070   & 25.48(30)  & 24.76(24) & 24.40(35)  & 23.73(24)    &  $>$21.8    &   $>$ 21.3   &   $>$20.1 \\
 21650   & 25.56(23)  & 25.75(34) & $>$ 24.9   & $>$ 24.5     &  $>$21.9    &   $>$ 21.4   &   $>$20.3 \\
2.44E6   & 25.85(24)  & 25.75(34) & $>$ 25.0   & $>$ 24.5     &  $>$22.2    &   $>$ 21.6   &   $>$20.6 \\
\bottomrule
\end{tabular}
\label{tab:logGROND.081226A}
\end{center}
\end{table}

\begin{table}[!h]
\caption[]{Log of the GROND observations of the afterglow of 
GRB 090305 (Fig.~\ref{fig:GROND090305lc}).}
\renewcommand{\tabcolsep}{1.5pt}
\begin{center}
\begin{tabular}{lllll}
\toprule
 Time (s) &$g^{\prime}$ &$r^{\prime}$ & $i^{\prime}$&$z^{\prime}$     \\
\midrule
  2014   &   --       & 23.13(09) &  22.96(18)  &       --         \\
  2568   & 23.77(19)  & 23.26(13) &   --        &       --         \\
  3318   & 23.79(15)  & 23.55(15) &  23.24(19)  &       --         \\
  3925   &   --       & 23.68(13) &   --        &       --         \\
  4367   &   --       & 23.61(07) &   --        &       --         \\
  4594   & 24.07(09)  &  --       &   23.55(13) &       --         \\
  4814   &  --        & 23.92(11) &   --        &       --         \\
  5262   &  --        & 23.82(20) &   --        &       --         \\
  5495   &  --        &  --       &   --        &       23.46(12)  \\
  5719   &  --        & 23.74(11) &   --        &       --         \\
  6166   &  --        & 23.91(08) &   --        &       --         \\
  6392   & 24.534(13) &   --      &  23.70(11)  &       --         \\
  6613   &  --        & 23.91(08) &   --        &       --         \\
  7065   &  --        & 23.86(08) &   --        &       --         \\
  7519   &  --        & 24.14(27) &   --        &       --         \\
\bottomrule
\end{tabular}
\label{tab:logGROND.090305}
\end{center}
\end{table}

\begin{table}[] 
\renewcommand{\tabcolsep}{3pt}
\begin{center}
\caption{Log of the Gemini observations of the afterglow of GRB 090305
(Fig.~\ref{fig:GROND090305lc}).}
\begin{tabular}{cc cc cc}
\toprule 
Mid-time   & $g'$           &  Mid-time      & $r'$              \\
(s)        & mag           &   (s)          & mag                \\
\midrule
   2859    & 23.89(05)     &   1681         &  23.18(03)         \\ 
   3329    & 23.94(05)     &   2150         &  23.21(03)         \\
   3800    & 24.02(07)     &   2621         &  23.43(04)         \\
    --     &  --           &   5220         &  23.77(05)         \\
    --     &  --           &   5689         &  23.82(05)         \\
    --     &  --           &   6159         &  23.89(05)         \\
    --     &  --           &   6478         &  24.04(04)         \\
    --     &  --           &   7587         &  24.29(04)         \\[1mm]
\bottomrule
\end{tabular}
\label{tab:logGemini.090305}
\end{center}
\end{table}

\begin{table}[!h]
\caption[]{Log of the GROND observations of the afterglow of 
GRB 090927 (Fig~\ref{fig:090927lc}).}
\renewcommand{\tabcolsep}{1.5pt}
\begin{center}
\begin{tabular}{lll}
\toprule
 Time (s) & $r^{\prime}$    & $i^{\prime}$             \\
\midrule
  61700   &    21.90(09)    &  21.79(06)              \\
  62380   &    21.86(15)    &   --                    \\
  63036   &    21.93(05)    &   --                    \\
  65325   &    21.89(07)    &   --                    \\
 150945   &    23.18(21)    &  23.03(22)              \\
\bottomrule
\end{tabular}
\label{tab:logGROND.090927}
\end{center}
\end{table}

\input{appendix1}
\input{appendix2}

\end{document}

%% file: Tab1.tex
\begin{table*}[htb]
\caption[]{The 20 short bursts of our sample.}
\renewcommand{\tabcolsep}{4pt}
\begin{center}
\begin{tabular}{rlll rcc lcl ll}
\toprule
\# &   GRB   & R.A. (J2000)  & Decl.      &Inst.& error [$''$] & Ref. &$T_{90}$ [s]     & Ref.            & $E(B-V)$ & $z$ & Ref.\\
(1)&   (2)   & (3)           & (4)        &(5)  & (6)          & (7)  &(8)              & (9)             & (10)     & (11)& (12)\\
\midrule
 1 & 070729  & 03:45:16.02         & $-$39:19:20.6        & XRT & 2.5       &   1       & 0.9$\pm$0.1     &   1   & 0.02  & --    & --  \\
 2 & 071112B & 17:20:51.0          & $-$80:53:02          & BAT & 132       &   2       & 0.3$\pm$0.05    &   2   & 0.12  & --    & --  \\
 3 & 071227  & 03:52:31.26         & $-$55:59:03.5        & OT  & 0.3       &   4       & 1.8$\pm$0.4     &   3   & 0.01  & 0.381 & 39 \\[1mm]
 4 & 080905A & 19:10:41.73         & $-$18:52:47.3        & OT  & 0.6       &  6,7      & 1.0 $\pm$ 0.1   &   5   & 0.14  & 0.122 & 41  \\
 5 & 080919  & 17:40:53.78         & $-$42:22:05.7        & XRT & 1.6       &  8        & 0.6$\pm$0.1     &   8   & 0.49  & --    & --   \\
 6 & 081226A & 08:02:00.45         & $-$69:01:49.5        & OT  & 0.2       &  this work       & 0.4$\pm$0.1     &   9   & 0.16  & --    & --   \\[1mm]
 7 & 081226B & 01:41:59            & $-$47:26:19          &IBIS & 150       &  11       & 0.7             &  11   & 0.02  & --    & --   \\   
 8 & 090305  & 16:07:07.59         & $-$31:33:21.9        & OT  & 0.2       & this work & 0.4$\pm$0.1     &  14   & 0.22  & --    & --   \\
 9 & 090426  & 12:36:18.07         &   +32:59:09.6        & OT & 0.5       &  this work & 1.2 $\pm$0.3     &  17   & 0.02  & 2.609 &  32  \\[1mm]
10 & 090510  & 22:14:12.50         & $-$26:34:59.0        &  OT & 0.2      &   42       & 0.3$\pm$0.1     &  18   & 0.02  & 0.903 &  33,43 \\      
11 & 090927  & 22:55:53.39         & $-$70:58:49.50       &  OT & 0.2     & this work   & 2.2$\pm$0.4     &  19   & 0.03  & 1.37  &  34 \\      
12 & 091109B & 07:30:56.61         & $-$54:05:22.85       &  OT & 0.5       &  20,38    & 0.3$\pm$0.03    &  21   & 0.03  & --    & --   \\[1mm]
13 & 091117A & 02:03:46.9          & $-$16:56:38          & BAT & 156       &  22,23    & 0.43$\pm$0.05   &  24   & 0.03  & --    & -- \\
14 & 100117A & 00:45:04.66         & $-$01:35:41.89       &  OT & 0.26      &  40      & 0.30$\pm$0.05   &  25   & 0.02  & 0.915 & 40 \\      
15 & 100206A & 03:08:39.03         &   +13:09:25.3        & XRT & 3.3       &  26       & 0.12$\pm$0.03   &  26   & 0.38  & 0.41  &  35  \\[1mm]
16 & 100625A & 01:03:10.91         & $-$39:05:18.4        & XRT & 1.8       &  27       & 0.33$\pm$0.03   &  27   & 0.01  & --    & --  \\
17 & 100628A & 15:03:52.41         & $-$31:39:30.2        & XRT & 7.0       &  28       & 0.036$\pm$0.009 &  28   & 0.17  & 0.102 &  36 \\    
18 & 100702A & 16:22:47.26         & $-$56:31:53.8        & XRT & 2.4       &  29       & 0.16$\pm$ 0.03  &  29   & 0.41  & --    & --  \\[1mm]
19 & 101129A & 10:23:41            & $-$17:38:42          & BAT & 180       &  30       & 0.35 $\pm$0.05  &  30 & 0.07  & --    & -- \\      
20 & 101219A & 04:58:20.49         & $-$02:32:23.0        & XRT & 1.7       &  31       & 0.6$\pm$0.2     &  32 & 0.06  & 0.718 & 37 \\[1mm]
\bottomrule
\end{tabular}
\label{tab:coords}
\end{center}
{\bf Notes:} \ 
The 5th, 6th, and 7th column give the instrument on which the coordinates are
based (OT stands for optical transient detected), the corresponding 
radius of the error circle and the reference, respectively. 
BAT and XRT stand for the instruments onboard of the \swift \
satellite, IBIS stands for the instrument onboard the \emph{INTEGRAL}
satellite. The 8th and 9th column provide $T_{90}$ and the 
corresponding reference. The last columns give the Galactic reddening 
$E(B-V)$ (mag) along the line of sight according to \citet{Schlegel1998}
as well as the redshift. 
If available, enhanced \swift/XRT positions are
given in columns \#3 and \#4 as well as  the revised error circles, taken from
http://www.swift.ac.uk/xrt$_-$positions/index.php
and \cite{Evans12250,Evans12273}.  
{\it References}: 
1  = \cite{GuidorziGCNR77},
2 = \cite{Perri103}, 
3  = \cite{Sakamoto2007GCN71561},
4  = \cite{DAvanzo7157},
5 = \cite{Pagani2008GCNR1621},
6  = \cite{Malesani8190}, 
7  = \cite{deUgarte8195}, 
8 = \cite{Preger2008GCNR172},
9  = \cite{Krimm2008GCN8735},
11  = \cite{Mereghetti2008GCN8734},
14  = \cite{Krimm2009GCN8936},
17  = \cite{Sato9263},
18 = \cite{Hoversten218},
19 = \cite{Grupe2009GCNR252},
20 = \cite{Levan10154},
21 = \cite{Oates259},
22 = \cite{Cummings2009GCN10171},
23 = \cite{D'Elia2009GCN10292},
24 = \cite{Sakamoto2009GCN10180},
25 = \cite{dePasquale2010GCNR269},
26 = \cite{KrimmGCNR271},
27 = \cite{HollandGCNR289},
28 = \cite{ImmlerGCNR290},
29 = \cite{Siegel2010GCNR292}.
30 = \cite{Cummings11436},
31 = \cite{Gelbord11461},
32 = \cite{Krimm11467},
32  = \cite{Levesque9264},
33  = \cite{Rau9353},
34  = \cite{Levan2009GCN9958},
35 = \cite{Cenko10389},
36 = \cite{Cenko10946},
37 = \cite{Chornock11518},
38 = \cite{Malesani10156},
39 = \cite{Avanzo2009},
40 = \cite{Fong2010},
41 = \cite{Rowlinson2010a},
42 = \cite{NicuesaGuelbenzu2012a},
43 = \cite{McBreen2010}.
\end{table*}

%% file: Tab2.tex
\begin{table*}[htb]
\renewcommand{\tabcolsep}{5pt}
\begin{center}
\caption{Summary of the 3$\sigma$ upper limits for the short-burst afterglows not
detected with GROND based on first-epoch data (AB magnitudes).}
\begin{tabular}{rl ll rlcc cccccccc}
\toprule 
\#   & GRB      &\ \ \ $t_{\rm GRB}$ & Ref.&  $t_{\rm start}^{\rm obs}\hspace*{1cm}$&\ \ \ mean  & $<dt>$      & $g'$  & $r'$ & $i'$ & $z'$ & $J$  & $H$  & $K_s$   \\
     &          & \ \ \ (UT)         &     &  (UT)        \hspace*{1cm}             &\ \ \ (UT)  & (hh:mm:ss)  &       &      &      &      &      &      &         \\
(1)  &    (2)   & (3)                & (4) &  (5)                                   &  (6)       &         (7) &  (8)  & (9)  & (10) & (11) & (12) & (13) & (14)  \\
\midrule
 1   & 070729   & 00:25:53     & 1   & 29-Jul-2007, 07:09:53& 09:13:33 & 08:47:40    & 24.5  & 24.7 & 24.4 & 24.3 & 22.7 & 21.8 & --    \\
 2   & 071112B  & 18:23:31     & 2   & 13-Nov-2007, 00:11:25& 00:49:15 & 06:25:44    & 24.6  & 24.4 & 23.8 & 23.5 & 21.6 & 20.7 & 20.0  \\
 3   & 071227   & 20:13:47     & 3   & 28-Dec-2007, 00:20:05& 00:23:14 & 04:09:27    & --    & 20.6 & 20.0 & 20.4 & 20.0 & 19.8 & 19.4  \\[1mm]
 4   & 080905A  & 11:58:54     & 4   & 06-Sep-2008, 05:22:14& 05:27:53 & 17:28:59    & 23.0  & 22.8 & 22.3 & 21.9 & 20.4 & 19.9 & 19.6  \\
 5   & 080919   & 00:05:13     & 5   & 19-Sep-2008, 00:13:31& 00:16:52 & 00:11:39    & --    & --   & --   & --   & 19.6 & 19.4 & 19.3  \\
     &          &              &     &              00:13:31& 00:20:34 & 00:15:21    & --    & --   & --   & --   & 19.8 & 19.5 & 19.5  \\[1mm]
     &          &              &     &              00:28:09& 00:39:16 & 00:34:03    & --    & --   & --   & --   & 19.8 & 19.7 & --    \\
     &          &              &     &              00:53:11& 01:22:37 & 01:17:24    & 23.5  & 22.7 & 22.2 & 21.9 & 19.7 & 19.7 & 19.8  \\      
 7   & 081226B  & 12:13:11     & 6   & 27-Dec-2008, 01:30:14& 02:00:55 & 13:47:44    & 25.5  & 25.2 & 24.3 & 23.9 & 22.0 & 21.5 & 20.5  \\
 12  & 091109B  & 21:49:03     & 7   & 10-Nov-2009, 03:31:40& 03:45:58 & 05:56:55    & 23.6  & 23.3 & 22.2 & 21.9 & 20.3 & 19.7 & 19.0  \\[1mm]
 13  & 091117A  & 17:44:29     & 8   & 19-Nov-2009, 00:46:48& 01:13:45 & 31:29:16    & 25.0  & 24.8 & 24.0 & 23.5 & 21.7 & 21.2 & 20.4  \\
 15  & 100206A  & 13:30:05     & 9   & 07-Feb-2010, 00:33:50& 01:09:43 & 11:39:28    & 24.7  & 24.4 & 23.9 & 23.1 & 21.7 & 21.3 & 20.4  \\[1mm]
 16  & 100625A  & 18:32:28     & 10  & 26-Jun-2010, 06:13:15& 06:43:04 & 12:10:36    & 23.6  & 23.1 & 22.8 & 22.9 & 21.8 & 21.2 & 20.3  \\
 17  & 100628A  & 08:16:40     & 11  & 29-Jun-2010, 01:24:19& 02:08:49 & 17:52:09    & 24.2  & 24.5 & 23.9 & 23.9 & 22.6 & 22.0 & 21.2  \\
 18  & 100702A  & 01:03:47     & 12  & 02-Jul-2010, 01:06:38& 01:09:51 & 00:06:04    & 24.1  & 23.6 & 23.0 & 22.5 & 20.4 & 20.0 & 19.3  \\[1mm]     
     &          &              &     &              01:06:38& 01:13:27 & 00:09:40    & 24.4  & 23.8 & 23.2 & 22.7 & 20.6 & 20.1 & 19.5  \\      
     &          &              &     &              01:21:04& 01:48:27 & 01:44:40    & 24.9  & 24.1 & 23.5 & 23.0 & 20.6 & 20.3 & 19.8  \\      
 19  & 101129A  & 15:39:32     & 13  & 30-Nov-2010, 06:20:25& 07:11:30 & 15:31:58    & 24.7  & 24.7 & 24.2 & 23.9 & 22.0 & 21.4 & 20.5  \\   
 20  & 101219A  & 02:31:30     & 14  & 19-Dec-2010, 03:55:06& 05:09:48 & 02:38:18    & 23.8  & 23.9 & 23.4 & 23.2 & 22.4 & 22.0 & 20.9  \\        
\bottomrule
\end{tabular}
\label{tab:ULs}
\end{center}
{\bf Notes:} \ 
Column \#3: GRB trigger time (UT); column \#5: time after the burst when the first optical OB\footnote{Observing block, a pre-define observing sequence} was started;
column \#6: mean observing time; 
column \#7: difference between column \#6 and \#3 (always in hh:mm:ss). 
columns \#8 -\#14: $3\sigma$ upper limits.
{\it Notes to individual bursts:}\
GRB 071227: just 1 OB was taken in the first night (4 min), the $g^\prime$ band is not useful;
GRB 080905A: just 1 OB was taken (8 min), $g^\prime r^\prime i^\prime z^\prime$ are calibrated based on GROND zeropoints;
GRB 081226B: The optical upper limits refer to the southern 50\% of the error
      circle, the other part was not imaged in $g^\prime r^\prime i^\prime z^\prime$;
GRB 100206A: This supercedes the information given in \cite{Nicuesa10396}.
{\it General:}\ Note that in the case of large BAT/IBIS error circles the true limiting
magnitudes of the afterglow might be 1 mag less deep than the
limiting magnitudes of the images as given here.
{\it References:} 
1  = \cite{GuidorziGCNR77},
2  = \cite{Perri103}, 
3  = \cite{Sakamoto7147},
4  = \cite{Pagani2008GCNR1621},
5  = \cite{Preger2008GCNR172},
6  = \cite{Mereghetti2008GCN8734},
7  = \cite{Grupe2009GCNR252},
8  = \cite{Cummings2009GCN10171},
9  = \cite{KrimmGCNR271},
10 = \cite{HollandGCNR289},
11 = \cite{ImmlerGCNR290},
12 = \cite{Siegel2010GCNR292},
13 = \cite{Cummings11436},
14 = \cite{Gelbord11461}.
\end{table*}

%% file: Tabgrbs.tex
\begin{table*}[t]
\caption[]{Summary of the data for the six bursts with optical afterglow 
detection by GROND.}
\renewcommand{\tabcolsep}{8pt}
\begin{center}
\begin{tabular}{lccc clr}
\toprule
 GRB    & $\alpha_1$    &  $\alpha_2$  &$t_b$ [ks]     &$\beta_{\rm opt}$ &$\Theta_{\rm jet} $[deg] & $E_{\rm cor}$ [erg]  \\
\midrule
081226A & $1.3\pm0.2$   &    --        & $>$10         &  --             & $\gtrsim$2.6& $\gtrsim2.1^{+1.3}_{-0.4}\,\times\,10^{47}$ \\
090305  &$0.56\pm0.04$  &$2.29\pm0.60$ & $6.6\pm0.4$   &$0.52\pm0.15$    & $2.2\pm0.2$ & $1.6^{+0.9}_{-0.4}\times10^{47}$  \\
090426  &$0.46\pm0.15$  &$2.43\pm0.19$ &$34.5\pm1.8$   &$0.76\pm0.14$    & $6.5\pm0.4$ & $4.2\pm1.4\,\times\,10^{48}$ \\
090510  &$1.13\pm0.11$  &$2.37\pm0.29$ &$1.6\pm0.4$    &$0.85\pm0.05$    & $\sim1$     & $\sim 3 \,\times\,10^{49}$          \\
090927  &$1.32\pm0.14$  &  --          & $>$600        &$0.57^{+0.17}_{-0.10}$& $\gtrsim12\pm2$ & $\gtrsim1.0^{+0.3}_{-0.2}\,\times\,10^{50}$ \\
100117A & $\sim1.3$     &   --         & $>$30         &  --             & $\gtrsim$2.4 & $\gtrsim 4.6\,\times\,10^{48}$ \\[1mm]
\bottomrule
\end{tabular}
\label{tab:summary}
\end{center}
{\bf Notes:} GRB 090426: The light-curve parameters refer to the wide jet
solution  (see paper I). GRB 090510: Light curve parameters of this burst are
interpreted as a jet at very early times. $\alpha_1$ as well as $t_b$ were
taken from the optical fit as reported in \cite{DePasquale2010}; $\Theta_{\rm
jet}$ and $E_{\rm cor}$ were taken from \cite{He2011}. For the other bursts
see this work. GRB 090927: Constraints on the jet break time come from the
X-ray data (Fig.~\ref{fig:090927lc}). The results refer to a wind model.
\end{table*}

%% file: appendix1.tex
\section{GRBs without afterglow detection by GROND}

\subsection{GRB 070729}

The original 90\% c.l. XRT error circle radius was 5\farcs7
(\citealt{Guidorzi6678}), which was refined to 4\farcs5 some hours later
(\citealt{Guidorzi6682}). A host galaxy candidate was soon reported
(\citealt{Berger6680}). However, the final XRT position lies  about 9$''$
north-east and does not overlap with the previous XRT error circle 
(\citealt{Evans12250,Evans12273}).

GRB 070729 was the first short GRB observed with GROND
after its commissioning in mid-2007.  GROND observations started 
6 hrs after the burst and continued  for 4.5 hrs until
sunrise. A second-epoch observation was performed the following night for 1
hr. No transient object between the two epochs was detected in any band 
(\citealt{Aybueke2008AIPC.1000}).  

\subsection{GRB 071227}

GROND started observing the field 4~hr after the GRB trigger.  At that time
the weather conditions were not good. GROND could not detect the afterglow in
any band (Table~\ref{tab:ULs}). Second-epoch observations were performed  the
following night. GROND was on target 29 hr after the burst and observed for
one hour. At that time the host galaxy had already been discovered by
\swift/UVOT (\citealt{Sakamoto7147,Cucchiara7150}), and its redshift was
measured to be $z = 0.381\pm0.001$
\citep{DAvanzo7152,Avanzo2009,Berger7154}. VLT observations revealed an
optical afterglow situated 3\farcs1 away from the centre of its host, an
edge-on galaxy \citep{DAvanzo7157,Avanzo2009}. GROND could not detect the
afterglow anymore,  only deep limiting magnitudes can be provided: $g^\prime
r^\prime i^\prime z^\prime JH$ = 25.5, 25.0, 24.2, 24.4, 21.5, 20.5 at 29 hrs
after the burst. The $r^\prime$-band upper limit is in agreement with the
expectations based on the VLT $R$-band detection at 0.3~d after the burst
if the optical afterglow was fading analogous to its X-ray counterpart with a
decay slope of $\alpha\sim$1 (see figure 7 in \citealt{Avanzo2009}). The GRB
host galaxy is discussed in detail by \cite{Avanzo2009}.

\subsection{GRB 080905A}

GROND started observing the field of GRB 080905A about 17.5 hrs after the
burst. Observations continued for only 11 min at a seeing of 2\farcs2.
The combined $g^\prime r^\prime i^\prime z^\prime$-band image as well as the
combined $JHK_s$-band image do not show the afterglow and faint host galaxy
discovered with the ESO/VLT (\citealt{Rowlinson2010a}). Our-non detection is
in agreement with these authors,  according to which at the time of our
observations the magnitude of the afterglow was around $R_C=24$, about 1 mag
below our detection limit.  Although the field is very crowded with stars, the
afterglow was situated in a region free of stars. In addition, it was well
separated from the center of its suspected anonymous host galaxy. Therefore,
the upper limits we can provide (Table~\ref{tab:ULs}) are not affected by the
light of the host galaxy. We refer to \cite{Rowlinson2010a} for a detailed
study of this burst and its host galaxy.

\subsection{GRB 091109B}

GROND observed GRB 091109B six hrs after the trigger. The weather conditions
over La Silla observatory were not good at that time.  Although GROND was on
target for one hour, observations were not deep enough due to clouds.  Inside
the 2\farcs8 90\% c.l. XRT error circle no source can be detected in the GROND
images (Table~\ref{tab:ULs}). 

A faint optical transient was discovered by VLT/FORS in the $R_C$ band at the
same time when GROND was observing (\citealt{Levan10154,Malesani10156}), but
it was not detected in the NIR (VLT/HAWKI). The non-detection of the afterglow
by GROND is in agreement with the magnitude reported by \cite{Levan10154},
$R_C\sim25$, which is deeper than our limiting $r^\prime$-band magnitude
(23.3; Table~\ref{tab:ULs}). Re-analysing the VLT/FORS data, we find that from
20~ks to 40~ks the light curve of the  afterglow can be fitted with a single
power law with a slope of $\alpha=0.80\pm0.04$. For this time period, there
are also simultaneous \swift/XRT observations which, within errors, can be
fitted with the same decay slope ($\alpha_{\rm X}=1.08\pm0.36$).

\subsection{GRB 100206A}

GROND started observing the field  11 hrs after the trigger. Observations were
performed at high airmass and under poor seeing conditions. No evidence for an
afterglow candidate was found in any band (\citealt{Nicuesa10396}).
\cite{Perley2011} published a detailed investigation of the GRB host galaxy. 

\subsection{GRB 100625A}

GROND visited the field of GRB 100625A several times. First-epoch observations
started 11.7 hrs after the GRB trigger and lasted for about 1
hour. Second-epoch observations were done on June 27, about 39 hrs after the
trigger, and a third run was performed on July 1 (about 5.5~d after the
trigger). Further data of the field were collected in 2010. 

Within the $r=1\farcs8$ 90\% c.l. XRT error circle (\citealt{Goad10886}) an
object is detected in all GROND epochs, the potential GRB host galaxy
(\citealt{Berger10897}).  No evidence was found in the GROND data for a
decaying  afterglow superimposed upon this galaxy (Table~\ref{tab:ULs}; note
that these upper limits refer to an isolated afterglow). 

\subsection{GRB 100628A \label{100628A.text}}

GROND started observing the field about 17 hrs after
the GRB trigger and remained on target for 1.5 hrs. At that time, two
extended objects were already detected inside the final 90\% c.l. XRT error
circle (\citealt{Berger10902, Berger10911}). No optical afterglow was detected.

\subsection{GRB 101219A}

Observations with GROND started about 80~min after the GRB trigger and
continued for about two hrs.  Although observations were performed under
good weather conditions (seeing 0\farcs8, airmass 1.1), the proximity of the
Moon  affected the depth of the observations.  No optical transient was
detected by GROND in any band  down to deep flux limits (Table~\ref{tab:ULs}).

\subsection{GRBs with arcmin-sized error circles}

This sample contains four bursts where only a \swift/BAT or, in one case, an
\emph{INTEGRAL}/IBIS error circle is known, which are typically 3
arcmin in radius. These events are GRBs 071112B, 081226B, 091117A, and
101129A.  Because of visibility constraints by GROND or
\swift/XRT in these cases GROND was on target not earlier than between 6 and
31 hrs after the corresponding GRB trigger. Given that, on average, short
GRB afterglows are intrinsically substantially fainter than those of long GRBs
(see \citealt{Kann2010,Kann2011}), it was not very likely that in these cases
GROND could image the afterglow in any band. Indeed, only upper limits can be
provided  (Table~\ref{tab:ULs}).

%% file: appendix2.tex
\section{Additional observations reported in the literature}

\subsection{GRB 070729}

\swift/BAT triggered on GRB 070729 at 00:25:53 UT (\citealt{Guidorzi6678}) and
had a duration of $T_{90}$(15-350 keV) $=0.9\pm0.1$~s
(\citealt{GuidorziGCNR77}). The burst was also seen by \emph{Konus A}
(\citealt{Golenetskii6690}). An uncatalogued X-ray source was found by
\swift/XRT but no optical afterglow by \swift/UVOT (\citealt{Guidorzi6678}).
Inside the initial $r=5\farcs7$ XRT error circle \cite{Berger6680} reported
the detection of  an extended object visible in the $K$ band. A refined XRT
error circle with a radius of $r=4\farcs5$ was later reported by
\cite{Guidorzi6682}. This error circle lies $3\farcs2$ away from the initial
XRT position. Optical follow-up observations were performed in the $R$ band
with the Swope 40-inch telescope at Las Campanas Observatory but no sources
were detected inside the XRT error circles, implying that the aforementioned
galaxy is a red object (\citealt{Berger6686}). No afterglow  was detected in
the radio band (\citealt{Chandra6742}). The position of the XRT afterglow was
later refined again and shifted by about 5$''$ in NE direction while it shrunk
to $r=2\farcs5$ (\citealt{Evans12250,Evans12273}).

\subsection{GRB 071227}

This was a bright and multi-peaked GRB with $T_{90}$(15-350 keV)
$=1.8\pm0.4$~s that triggered \swift/BAT at 20:13:47 UT
(\citealt{Sakamoto7147,Sato7148}).  It was also detected by \emph{Konus-Wind}
(\citealt{Golenetskii7155})  and \emph{Suzaku-WAM}
(\citealt{Onda7158}). \swift\ localized a bright X-ray afterglow
(\citealt{Beardmore7153}). UVOT observations
(\citealt{Sakamoto7147,Cucchiara7150}) revealed a single faint source near the
XRT error circle, which was identified as a galaxy also visible in the DSS
(\citealt{Berger7151}). VLT (\citealt{DAvanzo7152,Avanzo2009}) and Magellan
(\citealt{Berger7154}) spectroscopy revealed a redshift of this galaxy of $z =
0.381\pm0.001$ and further VLT follow-up detected the optical afterglow
(\citealt{DAvanzo7157,Avanzo2009}) at the tip of this edge-on spiral galaxy.

\subsection{GRB 080905A}

\swift/BAT and \fermi/GBM  triggered on GRB 080905 at 11:58:55 UT
(\citealt{Pagani8180,Bissaldi8204}). The BAT light curve shows 3 peaks with a
total duration of about 2 s (\citealt{Pagani8180}). Its duration was
$T_{90}$ (15-350 keV)=$1.1\pm0.1$s (\citealt{Cummings8187}).  A fading X-ray
afterglow was found but no optical afterglow was detected  with UVOT
(\citealt{Pagani8180}). A faint afterglow candidate was then discovered with
the VLT (\citealt{Malesani8190}) and also a host galaxy was seen
(\citealt{deUgarte8195}). The revised $r=1\farcs6$ XRT error circle is in
agreement with this afterglow position (\citealt{Evans8203}). The afterglow
is located in an outer arm of a star-forming spiral galaxy at $z = 0.1218
\pm 0.0003$, making it the closest short GRB known so far.  This event has
been analyzed in detail by \cite{Rowlinson2010a}.  

\subsection{GRB 080919}

GRB 080919 triggered \swift/BAT \ at 00:05:13 UT. The burst consists of a
single spike and had a total a duration of $T_{90}$ (15-350 keV)=$0.6\pm0.1$
s. \swift/XRT began observing about 71 s after the BAT trigger. The detected
X-ray afterglow could be localized with high precision ($r=2\farcs1$;
\citealt{Preger2008GCN8270}), but remained undetected already  from the second
orbit on. \swift/UVOT started observing about 11 s after XRT but no afterglow
candidate could be found in the {\it white} filter down to $m$=18
\citep{Preger2008GCNR172,Baumgartner2008GCN8275,Immler2008GCN8277}. The size
of the X-ray error circle could finally be improved to $r=2\farcs0$
(\citealt{Preger2008GCN8276}). Ground-based observations with the  robotic REM
telescope on La Silla, Chile, started already 74 s after the BAT trigger  and
revealed a bright NIR source in the XRT error circle ($H=13.73\pm0.03$), which
is also listed in the 2MASS catalog however and, therefore, might be an
unrelated Galactic foreground object \citep{Covino2008GCN8271}. No further
follow-up observations are reported in the literature.
The position of the XRT afterglow was
slightly refind three years after the event (\citealt{Evans12250,Evans12273}).

\subsection{GRB 081226A}

GRB 081226A triggered \fermi/GBM and \swift/BAT at 01:03:37 UT
\citep{Godet2008GCN8729,Kouveliotou2009GCN8785}. Its duration was
$T_{90}$(15-350 keV)$=0.4\pm0.1$~s \citep{Krimm2008GCN8735}. \swift/XRT
started observing the field 94.5 s after the BAT trigger, an afterglow was
found (\citealt{Godet2008GCN8738}). UVOT started observing 156 s after the
trigger but no optical afterglow was identified \citep{Hoversten2008GCN8737}.
Optical observations by ROTSE-IIIc starting 25 s after the GRB could only
reveal upper limits on any optical afterglow  (\citealt{Schaefer2008GCN8730}).
GROND detected an afterglow candidate (\citealt{Afonso2008GCN8731}),
but observations with Gemini-S did not reveal a fading behaviour,
neither of this source nor of a 2nd source found in the XRT error circle
(\citealt{Berger2008GCN8732, Berger2008GCN8736}).  No radio counterpart of the
optical afterglow candidate(s) could be found with the ATCA array in Australia
(\citealt{Moin2009GCN8952}). The position of the XRT afterglow was
slightly refind three years after the event (\citealt{Evans12250,Evans12273}).

\subsection{GRB 090305}

The burst triggered \swift/BAT at 05:19:51 UT. The BAT light curve shows a
single short spike with a duration of $T_{90}$(15-350 keV)$ =0.4\pm0.1$ s.
XRT began observing the field  93 s after the trigger, but no X-ray afterglow
was initially detected.  UVOT started observing 96 seconds after the trigger
but no optical afterglow candidate was discovered either
\citep{Beardmore2009GCN8932,Krimm2009GCN8936}. Despite the lack of an XRT
position, rapid follow-up of the BAT error circle with Gemini-S/GMOS
and Magellan/Baade led to
the discovery of the  optical afterglow \citep{Cenko2009GCN8933,
Berger2009GCN8934}. Furthermore, a re-analysis with relaxed constraints
allowed the detection of an extremely faint X-ray afterglow at the position of
the optical counterpart \citep{Beardmore2009GCN8937}. No host galaxy was
detected down to deep limits right under the optical afterglow position
\citep{Berger2010}.

\subsection{GRB 090927}

The burst was detected by \swift/BAT at 10:07:16 UT
(\citealt{Grupe2009GCN9945}). It had a FRED-like shape with some substructure
and a duration of  $T_{90}$(15-350 keV)$=2.2\pm0.4$ s
(\citealt{Stamatikos2009GCN9955}).  It was also detected by \fermi/GBM
\citep{Gruber2009GCN9974}. The final classification of the burst is not
totally clear. It is more likely a long GRB as it shows significant spectral
lag and was relatively soft \citep{Grupe2009GCNR252}. 
After its BAT trigger, \swift\ could not immediately  slew to the field
due to an Earth-limb constraint. When \swift/UVOT began observing the
field 2121 s after the trigger it immediately discovered an optical afterglow
candidate (\citealt{Gronwall2009GCN9946,Kuin2009GCN9954}).  Only thereafter
the detection of the X-ray afterglow was announced, a quite unusual situation
(\citealt{Evans2009GCN9950}). The afterglow was observed with the
1-m $f$/4 Zadko telescope in Western Australia (\citealt{Klotz2009GCN9956})
that started observations 50 minutes after the trigger (with the first
magnitude value for $t\sim$2 hr), with the Faulkes Telescope South in
Australia (\citealt{Cano2009GCN9960}) that observed 4.2 hrs after the onset
of the GRB, and with VLT/FORS2 on ESO Paranal \citep{Levan2009GCN9958} that
observed 16.5 hrs after the burst trigger. The VLT
observations allowed for a  measurement of the afterglow redshift ($z=1.37$;
\citealt{Levan2009GCN9958}). Radio observations with the Australian ATCA
array did  not reveal the afterglow \citep{Moin2009GCN10021}.   

\subsection{GRB 091109B}

\swift/BAT triggered on GRB 091109B at 21:49:03 UT   ($T_{90}$(15-350
keV)$=0.30\pm0.03$s; \citealt{Oates259}).  An X-ray afterglow was
immediately detected but no optical afterglow was found
(\citealt{Oates10148}). The burst was a symmetrical spike with no sign of
extended emission  (\citealt{Oates259}).  A faint, rapidly decaying afterglow
was discovered with the VLT at coordinates RA, Dec.(J2000) = 07:30:56.61,
$-$54:05:22.85  (\citealt{Levan10154,Malesani10156}).

\subsection{GRB 100117A}

The burst triggered \swift/BAT \citep{DePasquale2010GCN10336} and \fermi/GBM
\citep{Paciesas2010GCN10345} at 21:06:19 UT. It had a duration of
$T_{90}$(15-350 keV)$=0.3\pm0.05$ s \citep{Markwardt2010GCN10338}.  \swift/XRT
began observing the field 80 seconds after the BAT trigger and found a bright
X-ray afterglow which could be localized  with an uncertainty of $4\farcs6$
(radius) that could later be refined to $2\farcs4$
\citep{Sbarufatti2010GCN10342}. UVOT started observing about 1 min later but
could not find an optical counterpart \citep{DePasquale2010GCN10336,
DePasquale2010GCN10344}. The optical afterglow was detected by Gemini-North 8.3
hr after the burst with $r_{\rm AB} = 25.46\pm0.20$
\citep{Levan2010GCN10349}. The burst is in detail discussed in \cite{Fong2010}.

\subsection{GRB 100206A}

The burst triggered \swift/BAT at 13:30:05 UT (\citealt{Krimm10376}) and had a
duration of $T_{90} = 0.12\pm0.03$~s (\citealt{Sakamoto10379}). XRT started
observing the field 75~s after the trigger and found an uncatalogued X-ray
source (\citealt{Krimm10376}), whose coordinates were later refined to RA,
Dec.  (J2000)= 03:08:38.94, 13:09:25.5, with an error radius of $3\farcs2$
(\citealt{Goad10378}). The burst was also seen by \fermi/GBM with a spectral
peak at  $439^{+73}_{-60}$ keV, assuming a Band function
(\citealt{vonKienlin10381}). No  optical counterpart was detected by
\swift/UVOT  (\citealt{Krimm10376,Marshall10394}) and other ground-based
observatories  (\citealt{Bhattacharya10380, Guziy10384,
Noda10385,Leloudas10387,  Yurkov10391,  Mao10392, Andreev10455,
Rumyantsev10456}).  Evidence for a galaxy close to the XRT error circle was
soon reported  based on archival images of the field (\citealt{Miller10377}),
whose redshift was later determined to $z$=0.41
(\citealt{Cenko10389}). Morgan10390 found that at this  redshift this galaxy
is very bright in $JHK$, suggesting that this is a luminous infrared galaxy.
\cite{Levan10386} speculated about the discovery of the faint optical
afterglow  of GRB 100206A based on WHT observations starting 7 hr after the
burst.  However, no fading of this source was seen on Gemini images taken 7
and  11.5 hr after the event, suggesting that in fact this source could be the
true GRB host galaxy (\citealt{Berger10395,Berger10410}). This placed a limit
of $i>$24.7 on the brightness of the optical afterglow at 15.7 hrs after the
burst (\citealt{Berger10410}).  The position of the XRT afterglow was slightly
refind three years after the event (\citealt{Evans12250,Evans12273}).
\cite{Perley2011} dispute that another faint source  very close to the $z$ =
0.41 galaxy might instead be the host.

\subsection{GRB 100625A}

\swift/BAT triggered and located GRB 100625A at 18:32:28 UT
(\citealt{Holland10884}).  The BAT light curve showed a single spike with a
substructure and a duration of about 0.33 s. XRT started observing the field
48~s after the trigger and found an uncatalogued X-ray source, whose
coordinates were later refined to RA, Dec. (J2000)= 01:03:10.98,
$-$39:05:18.3, with an error radius of $1\farcs8$. No optical counterpart was
detected by \swift/UVOT. The burst was also seen by \emph{Konus-Wind}  and
\fermi.  The \fermi/GBM light curve shows two closely spaced narrow pulses
with a duration ($T_{90}$) of about 0.32 s (50-300 keV;
\citealt{HollandGCNR289}).  Inside the XRT error circle, an object was
reported in the optical bands by ground-based observatories
(\citealt{Levan10887, Berger10897, Tanvir10905}).  However, the non-variation
of the object and its extended shape pointed to it being a host galaxy
candidate.  The position of the XRT afterglow was slightly refind three years
after the event (\citealt{Evans12250,Evans12273}).

\subsection{GRB 100628A}

\swift/BAT triggered and located GRB 100628A  at 08:16:40 UT as did
\emph{INTEGRAL} (\citealt{Beckmann10898}). In the BAT window it had a duration
of $T_{90}$(15-350 keV)= $0.036\pm0.009$ s (\citealt{ImmlerGCNR290}).
\swift/XRT began observing the field  86 seconds later and located an X-ray
afterglow at coordinates RA, Dec.(J2000) = 15:03:52.95, $-$31:39:41.7, with an
error radius of $5\farcs2$ (\citealt{Starling10899}). No optical afterglow was
found, neither by UVOT nor by ground-based observatories
(\citealt{Immler10901,Burenin10900,Berger10902,Berger10911,Suzuki10904,
  Levan10909}). Based on Magellan observations, \cite{Berger10902} noticed
however the presence of several galaxies close to and inside the XRT error
circle (see also \citealt{Berger10908}). The position of the X-ray afterglow
was later rejected  when another faint X-ray source was found that had faded
away. This source at coordinates RA, Dec.(J2000)= 15:03:52.41, $-$31:39:30.2
(error radius 7$''$) is now considered as the most likely X-ray afterglow
(\citealt{Starling10941}). Inside this error circle \cite{Berger10943} reports
the presence of two galaxies, which however did not show any evidence  of a
superposed optical afterglow. For one of these galaxies \cite{Cenko10946}
measured a redshift of $z$=0.102.

\subsection{GRB 100702A}

The burst triggered \swift/BAT at 01:03:47 UT \citep{Siegel2010GCN10916}.  It
was a FRED-like single-peaked burst with  a duration of $T_{90}$(15-350
keV)$=0.16\pm0.03$ s \citep{Baumgartner2010GCN10926}.  \swift\ slewed
immediately to the burst and found a bright X-ray afterglow, which faded
rapidly after an early plateau phase  and was already undetected after the
first orbit \citep{Grupe2010GCN10924, Siegel2010GCNR292}.  No  optical/NIR
afterglow candidate was found, neither in rapid response observations by
ROTSE-IIIc located at Mt. Gamsberg, Namibia \citep{Flewelling2010GCN10917},
nor by optical observations with the acquisition camera of VLT/X-shooter
\citep{Malesani2010GCN10918} and NIR observations using  PANIC at the
Magellan/Baade telescope \citep{Fong2010GCN10919, Berger2010GCN10921}.

\subsection{GRB 101219A}

GRB 101219A was a short-hard burst localized by  \swift/BAT at 02:31:29 UT
\citep{Gelbord11461} and it was also detected by  \emph{Konus-Wind}
\citep{Golenetskii11470}.  The BAT light curve consists of a single spike with
a duration of $T_{90}$(15-350 keV)=$0.6\pm0.2$ s (\citealt{Krimm11467}). 
A fading X-ray afterglow was found 60~s  after the
trigger but no optical afterglow was detected \citep{Gelbord11461}. Inside the
XRT error circle a faint extended object was observed in the $i$ and $r$ band
with the Gemini South  telescope 43 min after the burst \citep{Perley11464}.
Furthermore, the same faint source was detected in the $J$ band with the 6.5-m
Magellan Baade telescope 1.5 hrs after the trigger \citep{Chornock11469}.
Second epoch observations showed no variable source inside the XRT error
circle. Spectroscopic observations performed on the host galaxy candidate with
GMOS  mounted at the Gemini-North telescope derived a redshift of $z$=0.718
\citep{Chornock11518}.  The position of the XRT afterglow was slightly revised
two years after the event (\citealt{Evans12250,Evans12273}).